\documentstyle[preprint,aps]{revtex}

\begin{document}
\tighten
\draft
\preprint{
 \parbox[t]{50mm}{hep-ph/9501280\\
 \parbox[t]{50mm}{CHIBA-EP-85\\
 \parbox[t]{50mm}{DPNU-94-51\\
}}}}

\title{Spontaneous Chiral-Symmetry Breaking in Three-Dimensional QED
with a Chern--Simons Term}
\author{K.--I. Kondo\cite{emkon}}
\address{Department of Physics, Chiba University,
Chiba 263, Japan}
\author{P. Maris\cite{emmar}}
\address{Department of Physics, Nagoya University,
Nagoya 464-01, Japan}
\date{December 16, 1994}
\maketitle
\begin{abstract}
In three-dimensional QED with a Chern--Simons term we study the phase
structure associated with chiral-symmetry breaking in the framework of
the Schwinger--Dyson equation.  We give detailed analyses on the
analytical and numerical solutions for the Schwinger--Dyson equation
of the fermion propagator, where the nonlocal gauge-fixing procedure
is adopted to avoid wave-function renormalization for the fermion. In
the absence of the Chern--Simons term, there exists a finite critical
number of four-component fermion flavors, at which a continuous
(infinite-order) chiral phase transition takes place and below which
the chiral symmetry is spontaneously broken.  In the presence of the
Chern--Simons term, we find that the spontaneous
chiral-symmetry-breaking transition continues to exist, but the type
of phase transition turns into a discontinuous first-order transition.
A simple stability argument is given based on the effective potential,
whose stationary point gives the solution of the Schwinger-Dyson
equation.
\end{abstract}
\pacs{11.30.Qc, 11.30.Rd, 11.10.Kk, 11.15.Tk}

\section{Introduction}

This paper is the detailed exposition of our previous letter
\cite{KM94} on chiral-symmetry breaking in (2+1)-dimensional QED with
$N$ flavors of four-component Dirac fermions (hereafter called QED3).
Such a model of QED3 is chirally symmetric in the absence of a bare
fermion mass term, $m_0 \bar \psi \psi$, in sharp contrast to the
(2+1)-dimensional gauge theory with two-component fermions, where we
cannot define chiral symmetry \cite{ABKW86,Pisarski84,SW88}.
Similarly to the four-dimensional case \cite{Miransky85}, the chiral
symmetry of QED3 may be broken spontaneously due to the dynamical
generation of a fermion mass. However the pattern of dynamical
symmetry breaking is shown to be qualitatively different from the
four-dimensional counterparts.

Recently, QED3 has found a vast region of application in condensed
matter physics as an effective theory in the long wavelength (or
low-energy) limit of a more realistic microscopic model
\cite{Fradkin91,Wilczek90,DM92}. Especially, since the discovery of
high-$T_c$ superconductivity and fractional quantum Hall effect, a
peculiarity of three-dimensional gauge theory, i.e., the existence of
a Chern--Simons (CS) term \cite{DJT81}
\begin{equation}
 {\theta \over 2} \epsilon^{\mu\nu\rho} A_\mu \partial_\nu
A_\rho
\end{equation}
has been a basic ingredient in these applications. The CS term gives
the gauge field a mass without destroying the gauge invariance.  And,
even if there is no bare CS term in the original Lagrangian, a CS term
can be generated by radiative corrections in three dimensions. In the
presence of a CS term, parity is broken explicitly, but the Lagrangian
still keeps the chiral symmetry when $m_0 \bar \psi \psi = 0$.  The
parity violation is a very important factor for the anyonic model to
be a candidate theory of high-$T_c$ superconductivity
\cite{Fradkin91,Wilczek90}.  A field-theoretic realization of such an
anyonic model consists of fermions interacting with an abelian
statistical gauge field whose dynamics is governed by a CS term. In
contrast to the chiral symmetry, it is known that the dynamical
breakdown of parity does not occur in QED3
\cite{ABKW86b,Poly88,RY86a,HM89,VW84,CCW91,KEIT94}.

In this paper we pay attention to the breaking of chiral symmetry in
the presence of a CS term within the framework of the Schwinger--Dyson
(SD) equation \cite{RW94}. In momentum space, the SD equation for the
full fermion propagator $S(p)$ is written as
\begin{equation}
 S^{-1}(p) = p_\mu \gamma_\mu - m_0 - e^2 \int {d^3q \over
(2\pi)^3}
 \gamma_\mu S(q) \Gamma_\nu(q,p) D_{\mu\nu}(k)\,,
\end{equation}
which should be solved self-consistently and simultaneously with other
SD equations for the full photon propagator $D_{\mu\nu}(k)$ and the
full vertex function $\Gamma_\mu(q,p)$. In order to solve these sets
of SD equations actually, however, we adopt some kind of
approximations as exemplified below.

There are several quantities we must specify to write down the closed
SD equation for the fermion propagator $S(p)$, namely, the full photon
propagator $D_{\mu\nu}(k)$ and the vertex function $\Gamma_\mu(p,q)$,
and we also have to choose a suitable gauge. In general, both the
photon propagator and the vertex satisfy their own SD equation, and
both depend on the gauge-fixing parameter $a$, if the covariant
gauge-fixing Lagrangian ${1 \over 2a}(\partial_\mu A^\mu)^2$ is taken.
There are also additional constraints on these $n$-point functions,
coming from relations such as the Ward--Takahashi (WT) identity
\begin{equation}
(p_\mu-q_\mu)\Gamma^\mu(p,q)=S^{-1}(p)-S^{-1}(q)\,.
\end{equation}
Now one of the problems in QED3 is to find a consistent truncation
scheme for the SD equation: as a consequence of the WT identity the
wave-function renormalization should be equal to the vertex
renormalization, which puts a constraint on the bare vertex
approximation. By using
\begin{equation}
 S(p) = [\gamma_\mu p_\mu A(p)   - B(p)]^{-1} \,,
\end{equation}
the SD equation for $S(p)$ is decomposed into two integral equations
for $A(p)$, the wave-function renormalization, and $B(p)$.  Although
with a bare vertex it is not possible to satisfy the WT identity
exactly, we should make sure to satisfy the constraint $A(p)\simeq 1$
when using the bare vertex approximation.

First of all, we restrict our attention to QED without CS term. In the
quenched case, in which the vacuum polarization to the photon
propagator is completely neglected, the simplest treatment is to
choose the Landau gauge $a=0$ and to take the bare vertex
$\Gamma_\mu(p,q) \equiv \gamma_\mu$. This quenched approximation is
equivalent to taking the $N \rightarrow 0$ limit in the SD equation
for the fermion propagator, since this limit eliminates the radiative
correction to the photon propagator coming from the internal fermion
loops (with $N$ species). Then the above SD equation leads to no
wave-function renormalization: $A(p) \equiv 1$, irrespective of $B(p)$,
see e.g. \cite{KN89}. In this case, it has been shown \cite{KN90} that
quenched QED3 resides in only one phase where chiral symmetry is
spontaneously broken, in agreement with the Monte Carlo simulation of
non-compact lattice QED3 \cite{DKK89}.

In the quenched case, the bare vertex approximation is justified in
the Landau gauge $a = 0$, because this choice is approximately
consistent with the WT identity and satisfies the requirement that the
wave-function renormalization and the vertex renormalization are
equal.  In gauges other than the Landau gauge however, we can not take
the bare vertex, since $A(p)$ deviates from one. In order to study the
SD equation in a general gauge, we must modify the vertex in such a
way that the result is gauge-covariant, i.e., independent of the
gauge-fixing parameter $a$.  This problem persists in the
three-dimensional case \cite{CPW92,AJM90} as well as in four
dimensions \cite{CP91,ABGPR93,Kondo92}. Such a modified vertex has to
satisfy the WT identity exactly.

Now we can raise the question whether the chiral symmetry restores at
a certain non-zero value of $N$.  In the presence of the one-loop
vacuum polarization ($N\not=0$) the simple procedure mentioned above
can not be applied, since there is no simple gauge choice $a$ such
that $A(p) \equiv 1$. Nevertheless, under the bare vertex
approximation in the Landau gauge, Appelquist, Nash and Wijewardhana
(ANW) \cite{ANW88} have shown that there is a finite critical number
of flavors $N_c$, above which the chiral symmetry restores. Based on a
leading-order $1/N$ expansion \cite{JT81} ($e^2= 8 \alpha/N$ with
$\alpha$ being kept fixed) for the vertex and the photon propagator in
the SD equation, they gave the critical value $N_c = 32/\pi^2 \cong
3.2$.  Furthermore, Nash \cite{Nash89} claimed that the leading-order
{\it gauge-invariant} critical number of flavors is given by
$N_c=128/3\pi^2$ and that, when $1/N^2$ corrections are included,
$N_c$ coincides approximately with this result.

However, such a simple treatment of the vertex function and the
wave-function renormalization function $A$ was criticised by
Pennington and Webb \cite{PW88} and Atkinson, Johnson and Pennington
\cite{AJP88}.  They claimed that, if the vertex is correctly improved,
using the WT identity, and the coupled equations for $A(p)$ and $B(p)$
are solved together, the finite critical number of flavors does not
exist. This implies that the chiral symmetry is spontaneously broken
in QED3 for all values of $N$, i.e., $N_c=\infty$. On the other hand,
the Monte Carlo simulation \cite{DKK89} of lattice non-compact QED3
seems to support a finite critical flavors:
\footnote{The $N \rightarrow \infty$ limit (with $\alpha$
being kept finite) corresponds to the weak-coupling limit
$\beta:=1/(\epsilon e^2(\epsilon)) \rightarrow \infty$ in the lattice
gauge theory defined on a lattice with lattice spacing $\epsilon$.}
$N_c \sim 3.5 \pm 0.5$, see also \cite{Azcoiti93}. However, it is
numerically very difficult to confirm the exponential decrease of the
dynamical mass for increasing $N$ found in \cite{PW88,AJP88}.

The origin of this controversy stems from the fact that in the
unquenched case $A(p) \equiv 1$ can not be deduced as a simple
consequence of the SD equation by choosing the Landau gauge, in sharp
contrast with the quenched case. Nakatani \cite{Nakatani88} has
proposed to use a nonlocal and momentum-dependent gauge function,
instead of the usual gauge-fixing term, in order to keep $A(p) \equiv
1$ and thus to overcome the inconsistency one has using the bare
vertex approximation in the Landau gauge. Actually the nonlocal gauge
found in \cite{GSC90,KM92,KEIT94} can play exactly the same role as
the Landau gauge in the quenched case, in the sense that $A(p) \equiv
1$ follows by choosing the appropriate nonlocal gauge in the bare
vertex approximation and that the nonlocal gauge $a(k)$ as a function
of the photon momentum $k$ reduces to the Landau gauge in the quenched
limit $N \rightarrow 0$. Taking into account only the leading and
next-to-leading order terms in the infrared region, which is the
essential region in QED3 \cite{ABKW86,KN92}, yields a finite critical
number of flavors \cite{KEIT94}
\begin{equation}
N_c = 128/3\pi^2 \cong 4.3 \,,
\end{equation}
which is the same as the leading-order result obtained by Nakatani,
and coincides with the result obtained by Nash in a different way
\cite{Nash89}. In this paper we use the nonlocal gauge as derived in
\cite{KEIT94}.

In the present paper the existence of a finite critical number of
flavors is confirmed by solving numerically the nonlinear SD equation,
using a bare vertex and the nonlocal gauge as derived in
\cite{KEIT94} in order to keep $A(p) \equiv 1$, without further
approximation. The parity-conserving and chiral-symmetry-breaking
fermion mass is dynamically generated in QED3 and there occurs a
chiral-symmetry-restoring phase transition at $N_c = 128/(3\pi^2)$.
The order of the chiral phase transition is infinite in the sense that
the dynamical fermion mass $m_d$ and the chiral order parameter
$\langle \bar \psi\psi\rangle$ exhibit an essential singularity at the
critical point $N=N_c$,
\begin{equation}
{\langle \bar \psi \psi \rangle \over \alpha^2}
\sim   \left(  {m_d \over \alpha} \right)^{3/2}
  \sim  \exp \left[ - {3\pi \over \sqrt{N_c/N-1}} \right]\,.
\end{equation}
\par
In the presence of CS term, as already reported in \cite{KM94}, the
chiral phase transition turns into a first-order transition, in sharp
contrast to the $\theta=0$ case. This result is obtained by both a
numerical study of the full SD equation and an analytical study of
approximated equations using the same scheme as in the absence of CS
term. We have also determined the critical line $N=N_c(\theta)$ for
this transition in the $(N,\theta)$ plane, the two-dimensional phase
diagram. In this paper we give the detailed exposition of this result
\cite{KM94} based on the framework of \cite{KEIT94}. The first-order
transition implies that the chiral order parameter as well as the
dynamical fermion mass show a discontinuous change at the critical
point on the whole critical line extending from $(N,\theta)=(N_c,0)$
in the phase diagram $(N,\theta)$.  Moreover, the critical number of
flavors $N_c(\theta)$ decreases as the CS coefficient $\theta$
increases.

This paper is organized as follows.  In section \ref{two} we give a
description of QED3 with a CS term. We start from the formulation of
QED3 using four-component fermions in the reducible representation of
the Clifford algebra for the $\gamma$-matrices in (2+1)-dimensions. We
define the chiral and the parity transformations for this theory, and
our decomposition of the fermion propagator into scalar functions. We
also discuss the structure of the gauge boson propagator with the
nonlocal gauge and give an explicit expression for the vacuum
polarization in the leading order of $1/N$ expansion. After these
preliminaries, we first discuss QED without the CS term and
subsequently study the effect of the explicit CS term. In section
\ref{secpure} we write down the SD equation for the fermion propagator
explicitly. The SD equation is solved both analytically and
numerically. Next, in section \ref{explcs}, we study the effect of the
explicit CS term by solving the nonlinear SD equation numerically. In
section \ref{csanal} detailed comparison of numerical and analytical
result is given and the numerical results in section \ref{explcs} are
confirmed by the analytical treatment. In order to study the stability
of the chiral-symmetry-breaking solution, we evaluate the effective
potential of Cornwall--Jackiw--Tomboulis \cite{CJT74} in section
\ref{CJTefpot}. The final section is devoted to conclusion and
discussion. In Appendix A, we give integration formulae which are
necessary to perform the angular integration to obtain the integration
kernel in the SD equation.  In Appendix B, we supplement details on
the calculation of the ultraviolet boundary condition. In Appendix C,
it is shown that at the stationary point the effective potential
obtained in section \ref{CJTefpot} actually gives the SD equation for
the fermion mass function in the nonlocal gauge.

\section{QED3 with Chern--Simons Term}
\label{two}

In Euclidean space the Lagrangian is
\begin{eqnarray}
{\cal L} &=& \bar\psi( i \not{\!\partial} + e\not{\!\!A}
        + m_e + \tau m_o )\psi
        + {\textstyle{1\over4}} F_{\mu\nu}^2
        - {\textstyle{1\over2}} i \theta \epsilon_{\mu\nu\rho}
                    A_\mu \partial_\nu A_\rho
        + {\cal L}_{\hbox{\scriptsize gauge fixing}} \,.
\end{eqnarray}
We use four-component spinors for the fermions, and accordingly a
four-dimensional representation for the $\gamma$-matrices of the
Clifford algebra $\{ \gamma_\mu, \gamma_\nu \}=-2\delta_{\mu\nu}$:
\begin{eqnarray}
 \gamma_0 \;\equiv\; \left( \matrix{ -i \sigma^3 & 0           \cr
                            0          & i \sigma^3 \cr }  \right)\,,
\;\;\;
\gamma_1 \;\equiv\; \left( \matrix{ i \sigma^1 & 0           \cr
                            0          & -i \sigma^1 \cr } \right)\,,
\;\;\;
\gamma_2 \;\equiv\; \left( \matrix{ i \sigma^2 & 0           \cr
                            0          & -i \sigma^2 \cr } \right)\,,
\end{eqnarray}
with $\sigma_a (a=1,2,3)$ being the Pauli matrices; furthermore we have
the matrix $\tau$, defined by
\begin{eqnarray}
 \tau = \left( \matrix{ 1 &  0 \cr
                        0 & -1 \cr } \right)\,.
\end{eqnarray}

With such a representation we can define chirality just as in
four-dimensional QED, but now there are two independent chiral
transformations possible, which are defined by the $4 \times 4$
matrices $\gamma_3$ and $\gamma_5$:
\begin{eqnarray}
 \gamma_3 \;\equiv\; \left( \matrix{ 0 & 1 \cr
                            1 & 0 \cr } \right)\,,
\;\;\;
 \gamma_5 \;\equiv\; \gamma_0 \gamma_1 \gamma_ 2 \gamma_3
 = \left( \matrix{ 0 & -1 \cr
                   1 &  0 \cr } \right) \,.
\end{eqnarray}
Without an explicit mass $m_e$ for the fermions, the Lagrangian is
chirally symmetric, but the mass term $m_e\bar\psi\psi$
breaks chiral symmetry. Note that the other mass term,
$m_o\bar\psi\tau\psi$, is chirally invariant.

In this representation, the parity transformation is defined by
$\psi(x_0,x_1,x_2) \rightarrow P \psi(x_0,-x_1,x_2)$,
$A_\mu(x_0,x_1,x_2) \rightarrow (-1)^{\delta_{\mu1 }}
A_\mu(x_0,-x_1,x_2)$, with $P=-i\gamma_5 \gamma_1$. Then the mass term
$m_o\bar\psi\tau\psi$ is odd under a parity transformation. Also the
CS term is odd under this parity transformation, so in the presence of
a (bare) CS term parity is always broken, even if $m_o = 0$. The other
terms in the Lagrangian, including the chiral-symmetry-breaking mass
term, are invariant under a parity transformation.

\subsection{The Fermion Propagator}
\label{twosubone}

The inverse full fermion propagator can be written as
\begin{eqnarray}
 S^{-1}(p) &=&
 A_e(p) \not{\!p} + A_o(p)\tau \not{\!p} - B_e(p) - B_o(p)\tau \,.
\end{eqnarray}
The functions $A(p)$ and $B(p)$ are scalar functions of the absolute
values of the momenta, and their bare values are $A_e = 1$, $A_0 = 0$,
$B_e = m_e$, and $B_o = m_o$. In order to study the fermion propagator
nonperturbatively it is useful to define the matrix (projection
operator)
\begin{eqnarray}
\chi_\pm = \frac{1}{2} ( 1\pm\tau) \,,
\end{eqnarray}
which allows us to rewrite the full propagator as
\begin{eqnarray}
 S(p) &=&  S_+(p) \chi_+ + S_-(p)\chi_- \nonumber\\
&=&
- \frac{A_+(p) \not{\!p} + B_+(p)}{A_+^2(p)p^2 + B_+^2(p)} \chi_+
- \frac{A_-(p) \not{\!p} + B_-(p)}{A_-^2(p)p^2 + B_-^2(p)} \chi_- \;,
\end{eqnarray}
where we have defined
\begin{equation}
 A_\pm = A_e \pm A_o \,,
\end{equation}
and
\begin{equation}
 B_\pm = B_e \pm B_o \,.
\end{equation}

In this paper we study the dynamical breaking of chiral symmetry,
using the SD equation for the fermion propagator. Perturbatively this
will not happen, but just as in pure QED (without CS term), chiral
symmetry can be broken dynamically due to nonperturbative effects,
starting with a chirally symmetric Lagrangian. So in the remainder we
have put both explicit masses $m_e$ and $m_o$ equal to zero
\footnote{We will reconsider this procedure in section VI.},
and study the behavior of $B_\pm$ nonperturbatively. Note that in
terms of $B_+$ and $B_-$, chiral symmetry means $B_+(p) = - B_-(p)$,
which gives $B_e(p) = 0$.

The order parameter connected with the chiral phase transition is the
chiral condensate. In the presence of a CS term there are two types of
condensates: a parity even condensate $\langle\bar\psi\psi\rangle$ and
a parity odd condensate $\langle\bar\psi\tau\psi\rangle$. Using the
decomposition of the propagator in terms of $\chi_\pm$, we can write
these condensates as
\begin{eqnarray}
   \langle\bar\psi\psi\rangle & = &
\langle\bar\psi\psi\rangle_+ + \langle\bar\psi\psi\rangle_-  \,,\\
   \langle\bar\psi\tau\psi\rangle & = &
\langle\bar\psi\psi\rangle_+ - \langle\bar\psi\psi\rangle_-  \,,
\end{eqnarray} where
\begin{eqnarray}
   \langle\bar\psi\psi\rangle_\pm & = & {1 \over \pi^2}
   \int_0^\infty {\rm d}k
    \frac{k^2\,B_\pm(k)}{A_\pm^2(k)k^2 + B_\pm^2(k)}    \,.
\end{eqnarray}

The general SD equation is given by
\begin{eqnarray}
 B_\pm(p^2) &=& e^2 \int\!\frac{{\rm d^3}k}{(2\pi)^3}
               \frac12 {\rm Tr}[\gamma_\mu S_\pm(k) \Gamma_\nu(p,k)
                D_{\mu\nu}(p-k)\chi_\pm]\,,\\
 A_\pm(p^2) &=& 1 + \frac{e^2}{p^2} \int\!\frac{{\rm d^3}k}{(2\pi)^3}
      \frac12 {\rm Tr}[{\,\not\!p} \gamma_\mu S_\pm(k) \Gamma_\nu(p,k)
        D_{\mu\nu}(p-k)\chi_\pm]\,.
\end{eqnarray}
In QED3, the usual truncation scheme for the fermion SD equation is
based on the $1/N$ expansion \cite{JT81}. The coupling constant $e^2$
has the dimension of mass, and we use the large $N$ limit in such a
way that $e^2 \downarrow 0$ and the product $N\,e^2$ remains fixed:
\begin{eqnarray}
 e^2 &=&  8 \alpha / N \,,
\end{eqnarray}
with $\alpha$ fixed. In this approximation scheme, the full vertex is
replaced by the bare vertex, because that is the leading-order
contribution in $1/N$.  In order to be consistent with the WT
identity, or at least with the requirement that the vertex
renormalization and the fermion wave-function renormalization are
equal, we use a suitable nonlocal gauge function.

\subsection{The Gauge Boson Propagator}
\label{twosubtwo}

We use a Lagrangian with a so called nonlocal gauge-fixing term for
the gauge field, namely
\begin{eqnarray}
{\cal L}(A;f(x-y)) & = &
        {\textstyle{1\over4}} F_{\mu\nu}^2
        - {\textstyle{1\over2}} i \theta \epsilon_{\mu\nu\rho}
                    A_\mu \partial_\nu A_\rho
        + \frac{1}{2}\int\!\!{\rm d}^3y\,f(x-y)
               \partial_\mu A_\mu(x) \partial_\nu A_\nu(y)  \,,
\end{eqnarray}
which has certain advantages above the normal gauge-fixing term: it
allows for a momentum-dependent gauge parameter in the gauge boson
propagator.

It is easy to show that this Lagrangian leads to the following inverse
bare photon propagator
\begin{eqnarray}
   D^0_{\mu\nu}{}^{-1}(q)
   & = & q^2 (\delta_{\mu\nu} - \frac{q_\mu q_\nu}{q^2})
        + \theta  \epsilon_{\mu\nu\rho} q_\rho
          + a^{-1}(q) q_\mu q_\nu   \,,
\end{eqnarray}
where we have defined the momentum-dependent gauge parameter in the
following way
\begin{eqnarray}
           a^{-1}(q) & = & \int\!\!{\rm d}^3x\,f(x) e^{-iqx}  \,.
\end{eqnarray}
So if we can simplify the actual calculations by choosing a specific
momentum-dependent gauge $a(q)$, we can justify this afterwards by
specifying the gauge-fixing term as
\begin{eqnarray}
  {\cal L} & = &   \frac{1}{2}\int\!\!{\rm d}^3y\,f(x-y)
       \partial_\mu A_\mu(x) \partial_\nu A_\nu(y)  \,,
\end{eqnarray}
with
\begin{eqnarray}
 f(x) & = &
\int\!\!\frac{{\rm d}^3q}{(2\pi)^3}  a^{-1}(q) e^{iqx}  \,.
\end{eqnarray}
The conventional (local) gauge can be recovered by the choice $f(x) =
{\delta(x)}/{a}$, as can easily be seen from the above formulae.

We can show that identities relating the different Green's functions,
like the WT identity, hold with this nonlocal gauge in exactly the
same way as with a constant gauge parameter. For the (full) photon
propagator this implies in momentum space
\begin{eqnarray}
   q^2 q^\mu D_{\mu\nu}(q)   & = & - a(q) q_\nu  \,.
\end{eqnarray}
This means that also with this nonlocal gauge the gauge-dependent part
of the photon propagator is not affected by the interactions, and that
the longitudinal part of the vacuum polarization is zero. Therefore we
can write the full photon propagator as
\begin{eqnarray}
 D_{\mu\nu}(q) & = &
 D^T(q^2)(\delta_{\mu\nu} - \frac{q_\mu q_\nu}{q^2})
                   + D^O(q^2)\epsilon_{\mu\nu\rho}\frac{q_\rho}{|q|}
                   + a(q) \frac{q_\mu q_\nu}{q^4} \,,\\
 D^T(q^2) &=& \frac{q^2 - \Pi^T(q)}
                     {(q^2-\Pi^T(q))^2 + (\Pi^O(q) - \theta |q|)^2} \,,\\
 D^O(q^2) &=& \frac{\Pi^O(q) - \theta |q|}
                     {(q^2-\Pi^T(q))^2 + (\Pi^O(q) - \theta |q|)^2} \,,
\end{eqnarray}
where $\Pi^T$ and $\Pi^O$ correspond to the decomposition of the
vacuum polarization tensor \cite{DJT81}:
\begin{eqnarray}
\Pi_{\mu \nu}(q) =   \left(
\delta_{\mu \nu}-{q_\mu q_\nu \over q^2} \right) \Pi^T(q)
 + \epsilon_{\mu \nu \rho} {q_\rho \over |q|} \Pi^O(q).
\end{eqnarray}

In the $1/N$ expansion the one-loop vacuum polarization has to be
taken into account, because this vacuum polarization is of order one:
there are $N$ fermion loops contributing to the vacuum polarization
and each loop is of the order $e^2 \sim 1/N$. Starting with massless
fermions, e.g. both the parity odd and parity even mass equal to zero,
there is no parity odd part of the vacuum polarization. The transverse
part of the vacuum polarization is
\begin{eqnarray}
  \Pi_T(q) &=& - \alpha |q| \,.
\end{eqnarray}
Therefore the inclusion of this vacuum polarization
leads to the following gauge boson propagator
\begin{eqnarray}
 D^T(q^2) &=& \frac{q^2 + \alpha |q|}
                        {q^2((|q| + \alpha)^2 +  \theta^2)} \,,\\
 D^O(q^2) &=& \frac{ - \theta |q|}
                        {q^2((|q| + \alpha)^2 +  \theta^2)} \,.
\end{eqnarray}
This is the photon propagator we will use in this paper, with a
suitable choice for $a(q)$ in order to keep the wave-function
renormalization equal to one.

\section{Dynamical Symmetry Breaking in Pure QED}
\label{secpure}

Without the CS term in the Lagrangian, we have no explicit parity
breaking terms, and the full fermion propagator will also be parity
even: there will be no spontaneous breaking of parity
\cite{ABKW86b,Poly88,RY86a,HM89,VW84,CCW91,KEIT94}. That means that we
only have to deal with one set of two coupled integral equations for
$A(p)$ and $B(p)$
\begin{eqnarray}
A(p^2) &=& 1 +
\frac{8 \alpha}{N \, p^2} \int\!\frac{{\rm d^3}k}{(2\pi)^{\rm 3}}
            \frac{A(k)}{k^2 A^2(k) + B^2(k)} \times  \nonumber\\ &&
\left(\left( D^T(q) - \frac{a(q)}{q^2}\right)\frac{2(p\cdot q)(k\cdot q)}{q^2}
     + \frac{a(q)\, p\cdot k}{q^2} \right) \,,\\
B(p^2) &=& \frac{8 \alpha}{N} \int\!\frac{{\rm d^3}k}{(2\pi)^{\rm 3}}
\frac{B(k)}{k^2 A^2(k) + B^2(k)}  \left(2 D^T(q) + \frac{a(q)}{q^2} \right) \,,
\end{eqnarray}
with $q = k - p $.

The condition that the wave-function renormalization is equal to one
leads to \cite{KEIT94}
\begin{eqnarray}  \label{gaugepure}
a(q) & = & 2 q^2 D^T(q) + {{4\,\alpha }\over |q|} - 2
 - {{4\,{\alpha^2}\over {{q^2}}}
 \,\ln \left({{\alpha  + |q|}\over {\alpha }}\right)} \,.
\end{eqnarray}
The SD equation for the fermions thus reduces to only one nonlinear
integral equation for the dynamical mass function, which we call
$m(p)$. After the angular integration, which can be done analytically
(see appendix A), the radial integration kernel becomes
\begin{eqnarray}  \label{kernpure}
{\rm K}(p,k) & = &  {{2\,\alpha }\over
  {{\rm max}(k\,,p)\,|k^2 - p^2|}}
+ {{1 }\over {k\,p}} \,\ln{\left(\frac{\alpha + |k+p|}{\alpha + |k-p|}\right)}
\nonumber\\ &&
 - {{{{\alpha }^2}\over {k\,p\,{{\left( k - p \right) }^2}}}
  \,\ln \left(1 + {{{|k - p |}}\over {\alpha }}\right)}
 + {{{{\alpha }^2}\over {k\,p\,{{\left( k + p \right) }^2}}}
  \,\ln \left(1 + {{{|k + p |}}\over {\alpha }}\right)} \,,
\end{eqnarray}
and the equation for the mass function is
\begin{eqnarray}  \label{massintpure}
 m(p) &=&  \frac{4\alpha}{\pi^2 N}\int_0^\infty\!\!{\rm d}k\,
      \frac{k^2\,m(k)}{k^2 + m^2(k)} {\rm K}(p,k) \,.
\end{eqnarray}
This nonlinear integral equation can be solved numerically without
further approximations, or it can be solved approximately by using a
series expansion for the logarithms. This last method makes it
possible to convert the integral equation into a second-order
differential equation, and to study dynamical chiral-symmetry breaking
analytically.

\subsection{Differential Equation}

In order to solve the equation analytically we make the replacement
$k^2 + m^2(k) \rightarrow k^2 + m^2(0)$, which is a good approximation
both for small momenta (where $m(k)$ is almost constant), and for
relatively large momenta (where both $m(k)^2$ and $m(0)^2$ are
negligible w.r.t. $k^2$). One can also show that this replacement is
in agreement with bifurcation analysis, see e.g.
\cite{Atkinson,Kondo92}. For very large momenta ($k > \alpha$) we
neglect $m^2$ with respect to $k^2$ in the denominator completely.

Another approximation, which is commonly made \cite{ABKW86}, uses the
fact that the integral is heavily damped for momenta larger than the
mass scale $\alpha$. All the essential physics comes from the infrared
part of the nonlinear interal equation. Therefore one uses a series
expansion for the logarithms in Eq.~(\ref{kernpure}) for momenta $ p,k
\ll \alpha$, and introduce a cutoff at $k = \alpha$. However, we are
interested in the behavior of the mass function for momenta $p >
\alpha$ as well, especially when we include the CS term.  Therefore we
will adopt a slightly different approximation, which takes into
account the ultraviolet tail of the integral as well.

For this purpose we expand the integration kernel ${\rm K}(p,k)$
in powers of $\min(p,k)/\max(p,k)$. To leading order this gives
\begin{eqnarray}
{\rm K}(p,k) & = &
2 \frac{\max(p,k)^2 + 2 \alpha^2
+ \alpha \max(p,k)}{\max(p,k)^3(\alpha+\max(p,k))}
    - \frac{4 \alpha^2}{\max(p,k)^4} \ln{\Big(1 + \max(p,k)/\alpha\Big)} \,.
\end{eqnarray}
Differentiating Eq.~(\ref{massintpure}) with this kernel leads to a
second-order differential equation
\begin{eqnarray}
  f(p) m''(p) + g(p) m'(p) & = &
             \frac{4\alpha}{\pi^2 N} \frac{p^2 m(p)}{p^2 + m^2(0)} \,,
\end{eqnarray}
with
\begin{eqnarray}
\label{purefp}
  f(p) & = & \frac{1}{K'(p,0)}               \,,\\
\label{puregp}
  g(p) & = & \frac{-K''(p,0)}{(K'(p,0))^2}   \,,
\end{eqnarray}
and two boundary conditions, infrared and ultraviolet ones, just as in
the usual approximation scheme.  It is easy to show that the behavior
of this differential equation in the infrared region is exactly the
same as one would obtain by expanding $f$ and $g$ for small momenta
directly.

For momenta $p \gg \alpha$ we expand the functions $f(p)$ and $g(p)$,
which gives to leading order in $p$
\begin{eqnarray}
 p^3 m''(p) +3 p^2 m'(p) + \frac{16\alpha}{\pi^2 N} m(p) &=& 0  \,.
\end{eqnarray}
To solve this we substitute a solution
\begin{eqnarray}
m(p) &=& p^a\sum_{i=0}^\infty c_i p^{-i}  \,,
\end{eqnarray}
and again taking into account leading order only we find $a = 0$ or $a
= -2$. It is easy to see that $a = 0$ is the solution corresponding to
explicit chiral-symmetry breaking, since this leads to $m(p)
\rightarrow m_0 \not= 0$ in the ultraviolet region. Without a bare mass,
the correct solution is $a = -2$, which shows that the dynamical mass
function falls off very rapidly in the ultraviolet region. This is
also consistent with the ultraviolet boundary condition, but it does
not provide a normalization condition.

With this knowledge we now consider integral equation in the infrared
region, {\em without neglecting the ultraviolet tail of the integral}:
we use
\begin{eqnarray}
 m(p) &=& \alpha^2 {m(\alpha) \over p^2}
\label{tail}
\end{eqnarray}
for $p > \alpha$, based on the ultraviolet behavior we have just
found, and normalized in such a way that the solution is continuous at
$p = \alpha$ (we do not require the continuity of derivative, $m'(p)$,
at $p = \alpha$). Using the usual approximation \cite{ABKW86} for
$p < \alpha$, we arrive at the integral equation
\begin{eqnarray}
 m(p) &=&  \frac{32}{3 \pi^2 N}\int_0^\alpha\!\!{\rm d}k\,
    \frac{k^2 \, m(k)}{k^2 + m^2(0)}{{1}\over {{\rm max}(k,p)}}
    + \frac{8\alpha}{\pi^2 N}\int_\alpha^\infty\!\!{\rm d}k\,
    \frac{m(k)}{k^2}  \,,
\end{eqnarray}
which finally reduces to
\begin{eqnarray}  \label{massexppure}
 m(p) &=&  \frac{32}{3 \pi^2 N}\int_0^\alpha\!\!{\rm d}k\,
    \frac{k^2 \, m(k)}{k^2 + m^2(0)}{{1}\over {{\rm max}(k,p)}}
    + \frac{8 m(\alpha)}{3 \pi^2 N}  \,.
\end{eqnarray}
This leads to exactly the same second-order differential equation in
the infrared region
\begin{eqnarray}  \label{diffmpure}
  p^2 m''(p) + 2 p m'(p)
  + \frac{32}{3\pi^2 N}\frac{p^2 m(p)}{p^2 + m^2(0)} &=& 0 \,,
\end{eqnarray}
as that obtained by neglecting the ultraviolet tail completely,
with also the same infrared boundary condition
\begin{eqnarray}
   m'(0) &=& 0  \,.
\end{eqnarray}

The general solution of Eq.~(\ref{diffmpure}), satisfying the infrared
boundary condition, is given by
\begin{eqnarray}
  m(p) & = & m(0)\,{}_2F_1(a_+,a_-,\frac32,-p^2/m(0)^2) \,,
\end{eqnarray}
with $a_\pm = \frac{1}{4} \pm \frac{1}{4}i\sqrt{N_c/N - 1}$ and a
critical number of fermion flavors
\begin{eqnarray}
  N_c &=& \frac{128}{3\pi^2}  \,,
\end{eqnarray}
above which there is no chiral-symmetry breaking. The only difference
is the ``ultraviolet'' boundary condition at $p = \alpha$. This
boundary condition now becomes
\begin{eqnarray}  \label{uvbcpure}
 m(\alpha) + \alpha m'(\alpha) &=& \frac{8 m(\alpha)}{3 \pi^2 N} \,,
\end{eqnarray}
due to the inclusion of the ultraviolet tail of the integral. This
boundary condition leads to a slightly different normalization of the
infrared mass $m(0)$ (see appendix B for more details)
\begin{eqnarray}
 \frac{m(0)}{\alpha} &=& \exp{\frac{-2\pi+ 2\phi}{\sqrt{N_c/N - 1}} } \,,
\end{eqnarray}
with
\begin{eqnarray}
 \phi & = &
\arg{\left[{\Gamma(1 + \frac{i}{2} \sqrt{N_c/N - 1})} {\Gamma(a_-)^2}
               \left({1 - \frac{1}{2}a_-}\right)\right]} \,.
\end{eqnarray}
Close to the critical number of flavors this can be expanded to give
\begin{eqnarray}
  m(0)/\alpha & = & \exp\left[ {\frac{-2\pi}{\sqrt{N_c/ N - 1}}
               + 3\ln{2} + \frac12\pi + \frac27} \right] \,,
\end{eqnarray}
which is almost the same as if one neglects the ultraviolet tail, in
which case the term $\frac{2}{7}$ in the exponent is absent.

However, this is not the only solution for the boundary condition;
it is known that there are infinitely many solutions if $N < N_c$
which behave in general as
\begin{eqnarray}  \label{pureosc}
  m(0)/\alpha & = & \exp\left[ {\frac{-2\, n\, \pi}{\sqrt{N_c/ N - 1}}
               + 3\ln{2} + \frac12\pi + \frac27} \right] \,,
\end{eqnarray}
close to the critical number of fermion flavors. The solution with the
largest value of $m/\alpha < 1$ corresponds to the ground state, since
this has the lowest energy. That means $n=1$ in the above equation,
and one can also show that the other solutions have an oscillating
behavior at large momenta; only the solution with $n=1$ is a nodeless
solution.

\subsection{Integral Equation}

Alternatively, we can solve the nonlinear integral equation for the
mass function, Eq.~(\ref{massintpure}), numerically, without further
approximations. The integral can be calculated numerically without any
cutoff, since the integrand falls off rapidly for large momenta and
the integral is finite. Solving the integral equation iteratively
leads to qualitatively the same result as the previous analysis.

There is a critical number of fermion flavors below which there is a
chiral-symmetry-breaking solution. The critical number is the same as
we have just found analytically
\begin{eqnarray}
  N_c & = & \frac{128}{3 \pi^2} \; = \; 4.32 \,,
\end{eqnarray}
and also the behavior of the infrared mass $m(0)$ is similar as
before.  In Fig.~\ref{fig1} we have plotted the infrared mass, or
actually $\sqrt{N_c/N - 1}\,\ln{(m(0)/\alpha)}$, versus
$\sqrt{N_c/N - 1}$ for the full integral equation together with the
analytical result. We can see that in both cases the mass behaves like
\begin{eqnarray}
  m(0)/\alpha & = &
\exp\left[ {\frac{-2\sigma_1}{\sqrt{N_c/ N - 1}} + \sigma_2} \right] \,,
\end{eqnarray}
the only difference is the value of the constants $\sigma_i$.  The
analytical result is $\sigma_1 = \pi$ and $\sigma_2 = 3.94$, whereas
the full nonlinear equation gives $\sigma_1 = 3.1$ and $\sigma_2 =
1.8$. So the main difference between the analytical solution of the
approximated equation and the numerical solution of the full equation
is an overall normalization factor. This difference is due to the fact
that in the first case we have made several approximations, but these
approximations turn out not to be essential for the behavior of the
infrared mass near the critical point. Both in the infrared and in the
ultraviolet region the numerical and analytical solutions have the
same behavior, see Fig.~\ref{fig2}. Of course, these approximations do
make a difference in the region where $p$ is of order $\alpha$, but
this is to be expected since we expand to leading order only in
$p/\alpha$ and $\alpha/p$ respectively.

\subsection{Chiral Condensate}

The order parameter of the chiral phase transition is the chiral
condensate
\begin{eqnarray}
   \langle\bar\psi\psi\rangle & = & {2 \over \pi^2}
   \int_0^\infty {\rm d}k
                  \frac{k^2 \, m(k)}{k^2 + m^2(k)} \,.
\end{eqnarray}
In order to get an explicit analytical formula for the condensate, we
can use Eq.~(\ref{massexppure}), which allows us to write the condensate
in terms of the mass function at $p = \alpha$. Using the same
linearization as before we get
\begin{eqnarray}
   \langle\bar\psi\psi\rangle & = & {2 \over \pi^2}
\left( \frac{3}{4} + \frac{3 \pi^2 N}{32} \right) \alpha m(\alpha)\,.
\end{eqnarray}
{}From this equation we can calculate how the chiral condensate behaves
close to the critical number of fermion flavors, see appendix B,
\begin{eqnarray}
  \frac{\langle\bar\psi\psi\rangle}{\alpha^2} & = &
  \frac{608}{7\,\pi^2} \frac{\Gamma{(1/2)}}{\Gamma{(1/4)^2}}
   \exp\left[ {\frac{-3\pi}{\sqrt{N_c/ N - 1}}
         + \frac{9}{2}\ln{2} + \frac34\pi + \frac37} \right] \, .
\end{eqnarray}
Alternatively, we can calculate the condensate numerically, using the
numerical solution of the full nonlinear integral equation, which
shows a similar behavior, see Fig.~\ref{fig1}.

\section{Explicit Chern--Simons Term}
\label{explcs}

In this section we add a CS term for the gauge field to the
Lagrangian. This breaks the parity explicitly, and gives rise to a
parity odd mass term for the fermions, as well as a parity odd part of
the gauge boson propagator.  We use the gauge boson propagator as
discussed in section \ref{twosubtwo}, with the inclusion of the
leading-order vacuum polarization.

Again we will use a nonlocal gauge-fixing term.  For the
momentum-dependent gauge function we use \cite{KEIT94}
\begin{eqnarray}
a(q) & = & 2 q^2 D^T(q) + 2 \Bigg( \frac{2\alpha}{|q|} - 1
  + \frac{4\alpha\theta}{q^2}
           \arctan{\frac{\theta |q|}{\alpha^2 + \alpha |q| + \theta^2}}
\nonumber \\ &&
  + \frac{\alpha^2 - \theta^2}{q^2}
           \ln{\frac{\alpha^2 + \theta^2}{(\alpha+|q|)^2 + \theta^2}}\Bigg) \,,
\end{eqnarray}
which leads to a wave-function renormalization almost equal to one, the
deviation of one is proportional to $\theta B_\pm$, which can be kept
very small.

\subsection{Schwinger--Dyson Equation}

With the above gauge boson propagator and gauge function, the
SD equation can be rewritten into two decoupled sets of
two coupled integral equations for $A_+$ and $B_+$, respectively for
$A_-$ and $B_-$. This leads to the following set of equations:
\begin{eqnarray}
\label{A}
A_\pm(p) &=& 1 \pm \frac{8 \alpha}{N\, p^2}
 \int\!\frac{{\rm d^3}k}{(2\pi)^{\rm 3}}
            \frac{2 B_\pm(k)}{k^2 A_\pm^2(k) + B_\pm^2(k)}
             D^O(q) \frac{p\cdot q}{|q|} \,,\\
\label{B}
B_\pm(p) &=& \frac{8\alpha}{N}\int\!\frac{{\rm d^3}k}{(2\pi)^{\rm 3}}
            \frac{1}{k^2 A_\pm^2(k) + B_\pm^2(k)} \times  \nonumber\\ &&
           \Bigg(B_\pm(k)\Big(2 D^T(q) + \frac{a(q)}{q^2} \Big)
  \mp 2 A_\pm(k) D^O(q) \frac{k\cdot q}{|q|}\Bigg) \,.
\end{eqnarray}
The even and odd parts of the scalar function $B(p)$ are
\begin{eqnarray}
  B_e(p) &=& (B_+(p) + B_-(p))/2 \,,
\end{eqnarray}
and
\begin{eqnarray}
  B_o(p) &=& (B_+(p) - B_-(p))/2 \,,
\end{eqnarray}
in terms of $B_\pm(p)$, and similar for $A(p)$. In analyzing these
equations it is important to observe that once we have found a
solution for $A_+(p)$ and $B_+(p)$, we {\em automatically} have also a
solution for $A_-(p)$ and $B_-(p)$: namely the set $A_-(p) = A_+(p)$
and $B_-(p) = -B_+(p)$. That means that we can always construct a
chirally symmetric (but parity odd) solution of the SD equation, with
$B_e(p) = 0$.  The question of dynamical chiral-symmetry breaking
turns into the question whether or not there exist {\em two} (or more)
solutions of the set of integral equations.

Without the CS term there is dynamical chiral-symmetry breaking only
for $N < N_c = 128/(3 \pi^2)$ as we have just seen. We therefore
expect a similar situation in the presence of the CS term, at least if
the parameter $\theta$ is small. That means that for $N > N_c$ we have
only the chirally symmetric solution of the above equations, but for
$N < N_c$ we expect that there are (at least) two solutions for both
$B_+$ and $B_-$ possible, which can be distinguished by their behavior
under the chiral and parity transformations and by their behavior in
the limit $\theta \downarrow 0$.

It is also essential to note that in the presence of an explicit CS
term for the gauge field in the Lagrangian there is no trivial
solution $B = 0$, as there would be without the explicit CS term. Due
to the explicit breaking of parity, the fermions always acquire a
parity-odd mass $B_o$, even if the explicit odd mass $m_o$ is zero. At
large momenta the CS term in the SD equation dominates (at least if
there is no explicit mass term present), which leads to an ultraviolet
behavior of the mass functions
\begin{eqnarray}
  B_+(p) \sim -B_-(p) \sim \frac{\theta}{p} \,,
\end{eqnarray}
at a perturbative level, whereas in the absence of the CS
term the mass function $m(p)$ falls off like $1/p^2$, as we have seen
in the previous section.  We will discuss this point in more detail
when we are studying the breakdown of chiral symmetry analytically.

\subsection{Numerical Results}

We have first solved the set of coupled integral equations
Eq.~(\ref{A}) and (\ref{B}) numerically without further
approximations. Depending on the values of the number of fermion
flavors $N$ and the CS coefficient $\theta$, there is only one
(chirally symmetric) solution in most parameter space, but we found
two solutions, allowing for a chiral-symmetry-breaking solution, for
small $\theta$ and small $N$, in agreement with the above
expectations. Using the notation $B_\pm(p)$ and $\tilde B_\pm(p)$ for
the two different solutions of the above equations, we have found the
following solutions:

\begin{enumerate}
\item
$B_+(0) = - B_-(0) = {\cal O}(m(0)) > 0$, with $B_+(p) = - B_-(p) > m(p) $, \\
$B_e(p) \equiv 0$ and $B_o(0) = {\cal O}(m(0))$        \\
for $N < N_c(\theta=0)$ we find that $B_\pm(p)$ and thus
$B_o(p)$ go towards the nontrivial solution $m(p)$ in the limit
$\theta\downarrow 0$; \\
for $N > N_c(\theta=0)$ we find that
$B_+(p) = B_o(p) = {\cal O}(\theta)$, which vanishes in the limit
$\theta\downarrow 0$.

\item
$B_+(0)  \sim \tilde B_-(0) = {\cal O}(m(0))$,
with $B_+(p) > m(p) > \tilde B_-(p) $  \\
$B_e(p) = {\cal O}(m(p))$ and $B_o(p) = {\cal O}(\theta)$                \\
This solution exists for values of $N < N_c(\theta = 0)$ and small
(compared to $\alpha$) values of $\theta$ only. There is a critical
number $N_c(\theta) < N_c(0)$ for given $\theta$, or critical
$\theta_c(N)$ for given $N < N_c(0)$. In the limit $\theta\downarrow
0$ both $B_\pm$ and thus $B_e$ go towards the nontrivial solution
$m(p)$: $B_+(p)\downarrow m(p)$ and $\tilde B_-(p) \uparrow m(p)$ in
this limit. The parity breaking solution $B_o(p)$ vanishes in this
limit.

\item
$\tilde B_+(0) = - \tilde B_-(0) = - {\cal O}(m(0)) < 0$ \\
The chirally symmetric combination, which exists for small values of
$N$ and small values of $ \theta$ only.

\item
$\tilde B_+(0)  \sim B_-(0) = - {\cal O}(m(0))$ \\
The fourth possible combination of $B_+$ and $B_-$, also existing for
small values of $N$ and small values of $ \theta$ only.
\end{enumerate}

The typical behavior of the numerical solutions for $A$ and $B$ is
shown in Fig.~\ref{fig3}. As expected, $A_\pm$ is indeed very close to
one, due to our choice of the gauge function. The solution $B_\pm(p)$
exists for all values of both $N$ and $\theta$, allowing only a
chirally symmetric solution. The other solution, $\tilde B_\pm(p)$, is
the interesting one, leading to dynamical chiral-symmetry breaking.
The iterative process of solving the integral equation numerically
does not converge to a (second) stable solution $\tilde B_\pm$ for all
values of $N$ and $\theta$: we can find this solution only for $N$
below some critical value (depending on $\theta$) and $\theta$ below
some critical value (depending on $N$), thus showing a chiral
phase transition at some critical $\theta$ and $N$.

Our numerical results all indicate strongly that this phase transition
is a {\em first-order} phase transition, in contrast to the
infinite-order phase transition in pure QED ($\theta = 0$), although
it is numerically very difficult to establish the type of phase
transition at the critical values of $N$ and $\theta$. For fixed
$\theta$ and small $N$, there exists a second solution, $\tilde
B_\pm(p)$, as can be seen from Fig.~\ref{fig4}. The value at the
origin $\tilde B_-(0)$ decreases rapidly for increasing $N$ and this
solution ``disappears'' (without $B_\pm(0)$ going to zero) at some
critical value $N < N_c(\theta=0)$, and also $B_e(0)$ does not go to
zero at this critical value. Considering $B_e(0)$ as the order
parameter for the chiral phase transition, this corresponds to a
first-order phase transition, in contrast to the infinite-order
phase transition at $\theta = 0$.

If we look at the behavior of $\tilde B_-(0)$ and $B_e(0)$ at fixed $N
< N_c(\theta=0)$ and increase $\theta$, we see a similar situation.
For (very) small values of $\theta$ we find the two solutions $B$ and
$\tilde B$ leading to chiral-symmetry breaking. For increasing
$\theta$ we find that $\tilde B(0)$ decreases, and disappears at some
critical value $\theta_c(N)$, without going to zero at this critical
value, see Fig.~\ref{fig5}.

\subsection{Condensate}

Although both functions $B_\pm(p)$ behave like $\theta/p$ in the
ultraviolet region, leading to divergent integrals for the condensates
$\langle\bar\psi\psi\rangle_\pm$, the chiral condensate is convergent
due to the fact that the leading-order contributions in $B_+(p)$ and
$B_-(p)$ (or $\tilde B_-(p)$) cancel. So the chiral order parameter is
\begin{eqnarray}
   \langle\bar\psi\psi\rangle & = & {1 \over \pi^2}
   \int_0^\infty {\rm d}k \left(
                  \frac{k^2\,B_+(k)}{A_+^2(k)k^2 + B_+^2(k)}
      +\frac{k^2\,\tilde B_-(k)}{\tilde A_-^2(k)k^2 + \tilde B_-^2(k)}
     \right) \,.
\end{eqnarray}
The chirally symmetric combination, $(B_+(p),B_-(p))$ with $B_e = 0$,
gives $\langle\bar\psi\psi\rangle = 0$, as would be expected from a
chirally symmetric solution.  Note that the other condensate,
$\langle\bar\psi\tau\psi\rangle$, is actually logarithmic divergent
because the leading-order contributions add up. Once we have the
numerical solutions, it is straightforward to calculate this chiral
condensate as well, see Fig.~\ref{fig5}. The behavior of the chiral
condensate also indicates a first-order phase transition.

However, as we mentioned before, it is numerically very difficult to
establish such a first-order phase transition. In order to confirm
that it is indeed a first-order transition and to determine the
critical parameters $N_c$ and $\theta_c$, we have to study the phase
transition analytically. We can do this by solving the integral
equations analytically, after some further approximations analogously
to the approximations leading to analytic solution in the pure QED
case.

\section{Analytical Study}
\label{csanal}

In order to see whether there is indeed a first-order phase
transition, we have analyzed the SD equation analytically, after some
more approximations. Based on the fact that $A(p) \equiv 1$ exactly if
$\theta$ is zero, and very close to one for small values of $\theta$
(small compared to $\alpha$), we replace $A(p)$ by one, so we get the
following equation for $B_\pm$
\begin{eqnarray}   \label{explmasseq}
B_\pm(p) &=& \frac{8\alpha}{N}\int\!\frac{{\rm d^3}k}{(2\pi)^{\rm 3}}
            \frac{1}{k^2 + B_\pm^2(k)} \times  \nonumber\\ &&
           \Bigg(B_\pm(q)\Big(2 D^T(q) + \frac{a(q)}{q^2} \Big)
  \mp 2 D^O(q) \frac{k\cdot q}{|q|}\Bigg) \,.
\end{eqnarray}

The error we make in neglecting the effects of the wave-function
renormalization (even if we use the nonlocal gauge) is of the order of
$\theta B_\pm$, which is (close to the critical number of fermion
flavors in the absence of the CS term) of the order of $\theta^2$,
because the parity odd mass, generated by the CS term, is of order
$\theta$. So for a consistent approximation in order to get an
analytical solution of the equation, it is enough to expand the kernel
in $\theta$ and neglect all terms of order $\theta^2$ and higher.
\footnote{This should be compared with the analysis by Hong and Park
\cite{HP93} where the order $\theta$ term was neglected from the
beginning.} Expanding the nonlocal gauge in $\theta$ gives
\begin{eqnarray}
a(q) & = &{2\,q\over {\alpha + q}} + {{4\,\alpha} \over q } - 2
  - {{4\,\alpha^2}\over {q^2}}
   \,\ln \left({{{\alpha + q }}\over {\alpha}}\right)
   + {{{\cal O}(\theta^2)}}  \,.
\end{eqnarray}
So to order $\theta$ it is the same as the nonlocal gauge without the
CS term, see Eq.~(\ref{gaugepure}). By inspection of the transverse part
$D^T(q^2)$ of the photon propagator it is easy to see that up to order
$\theta$ this is also exactly the same as without the CS term.  That
means that the second term on the RHS of Eq.~(\ref{explmasseq}),
proportional to $D^O(q)$, is the only ${\cal O}(\theta)$ contribution
in the SD equation for $B_\pm$, and that for the other terms we can
use the same kind of approximations as in section \ref{secpure} on
pure QED.

So in order to study the problem of dynamical chiral-symmetry breaking
analytically, we replace $k^2 + B^2_\pm(p)$ by $k^2 + M^2_\pm$, where
$M_\pm \equiv B_\pm(0)$, in the denominator of the integrand. In
general we get
\begin{eqnarray}
  B_\pm(p) & = & \frac{32}{3 \pi^2 N} \int_0^\infty{\rm d}k\,
                \frac{k^2 B_\pm(k)}{k^2 + M_\pm^2} K(p,k)
\nonumber\\&&
\pm \frac{4\alpha\theta}{\pi^2 N}
\int_0^\infty \frac{k^2 \, {\rm d}k}{k^2+M_\pm^2}
       \int_{-1}^1{\rm d}z \frac{k \cdot (k-p)}{(k-p)^2(|k-p| + \alpha)^2} \,,
\end{eqnarray}
with the kernel given by Eq.~(\ref{kernpure}), and $z = \cos{\phi}$,
with the angle $\phi$ between the vectors $p$ and $k$: $p\cdot k =
pk\,\cos{\phi}$.  For momenta $p < \alpha$ this reduces to
\begin{eqnarray}  \label{explappr}
  B_\pm(p) & = & \frac{32}{3 \pi^2 N} \int_0^\alpha{\rm d}k\,
                \frac{k^2}{k^2 + M_\pm^2}\frac{B_\pm(k)}{\max(p,k)}
    + \frac{8\alpha}{\pi^2 N}\int_\alpha^\infty\!\!{\rm d}k\,
    \frac{B_\pm(k)}{k^2}
\nonumber\\&&
\pm \frac{4\alpha\theta}{\pi^2 N}
\int_0^\infty \frac{k^2 \, {\rm d}k}{k^2+M_\pm^2}
       \int_{-1}^1{\rm d}z \frac{k \cdot (k-p)}{(k-p)^2(|k-p| + \alpha)^2} \,.
\end{eqnarray}
In order to get this equation, we have used the same approximations as
in section \ref{secpure} for the first term of the integral, which is
independent of $\theta$.

\subsection{Explicit Chern--Simons Contribution}

The last term on the RHS in Eq.~(\ref{explappr}), proportional to $\theta$,
\begin{eqnarray}
 I_\theta(p)&\equiv& \frac{4\alpha\theta}{\pi^2 N}
       \int_0^\infty \frac{k^2 \, {\rm d}k}{k^2+M_\pm^2}
       \int_{-1}^1{\rm d}z \frac{k \cdot (k-p)}{(k-p)^2(|k-p| + \alpha)^2} \,,
\end{eqnarray}
can be calculated analytically, see appendix A. For the radial
integration we expand the integration kernel for $p < q$ and $p > q$
and take into account only the leading-order terms in
$\min(p,q)/\max(p,q)$. Furthermore, we neglect $M_\pm$ with respect to
$\alpha$, which is justified close to the critical number of fermion
flavors and for small values of $\theta$. So in the far infrared
region the leading contribution coming from the explicit CS term
behaves like
\begin{eqnarray}
  I_\theta(p) & = & \frac{8}{\pi^2 N}\, \theta + {\cal O}{(p)} \,,
\end{eqnarray}
and in the far ultraviolet region the leading-order behavior is
\begin{eqnarray}
  I_\theta(p) & = &  \frac{32}{3\pi^2 N}\,\frac{\alpha}{p}\, \theta
                         + {\cal O}{(1/p^2)} \,.
\end{eqnarray}
This means that in the ultraviolet region the CS term will
dominate, since we know that the mass function $B(p)$ in the absence
of the CS term behaves like $1/p^2$ in the far ultraviolet.
Higher order contributions in $\min(p,q)/\max(p,q)$ will change this
result only quantitatively, but not affect the general behavior.

Based on these expansions, we use an analytical (continuous)
approximation for the contribution from the $\theta$ term
\begin{eqnarray}
  I_\theta(p) & = & \left\{ \begin{array}{lcl}
   \frac{8\theta}{\pi^2 N} && {\hbox{for $p < \alpha$}} \\
   \frac{32\theta}{3\pi^2 N}\left(\frac{\alpha}{p}
- \frac{\alpha^2}{4 p^2} \right)
                           && {\hbox{for $p > \alpha$}}
                     \end{array} \right.
\end{eqnarray}
for small values of $\theta$ compared to $\alpha$.

\subsection{Ultraviolet Behavior}

The next thing we have to calculate before we solve the integral
equation in the infrared regoin, is the ultraviolet tail
\begin{eqnarray}
 \frac{8\alpha}{\pi^2 N}\int_\alpha^\infty\!\!{\rm d}k\,
 \frac{B_\pm(k)}{k^2} \,,
\end{eqnarray}
but for this purpose we have to know the ultraviolet behavior of the
mass function. From the integral equation we can derive a second-order
differential equation for momenta $p \gg \alpha$
\begin{eqnarray}
 f(p) (B_\pm''(p) \mp I''_\theta(p)) + g(p) (B_\pm'(p) \mp I'_\theta(p)) & = &
             \frac{4\alpha}{\pi^2 N} B_\pm(p)  \,,
\end{eqnarray}
with $f(p)$ and $g(p)$ given by Eq.~(\ref{purefp}) and (\ref{puregp})
respectively.  In the ultraviolet region, these functions behave like
\begin{displaymath}
\begin{array}{rclrcl}
  f(p) &=& {\displaystyle\frac{- p^3}{4} }
\,, &  g(p)  &=& {\displaystyle\frac{-3 p^2}{4}}   \,,\\
\\
  I'_\theta(p) &=& {\displaystyle\frac{- 32\theta\alpha}{3\pi^2 N p^2}}
\,, \hspace{5mm}&
  I''_\theta(p) &=&
   {\displaystyle\frac{ 64\theta\alpha}{3\pi^2 N p^3}\,.}
\end{array}
\end{displaymath}
The leading ultraviolet behavior of the solution of this equation is
either constant (which would correspond to an explicit mass term in
the original integral equation), or
\begin{eqnarray}
 B_\pm(p)\, = \,\pm I_\theta(p) \,=\,
 \frac{32}{3\pi^2 N}\,\frac{\alpha}{p}\, \theta \,,
\end{eqnarray}
which is the correct solution in this case, and also in agreement with
the ultraviolet boundary condition. This is the same as the
perturbatively dominant behavior, as could also be seen directly from
the original integral equation, assuming that the mass function falls
off rapidly enough in the ultraviolet region. In order to have a
continuous solution at $p = \alpha$, we use the next-to-leading order
term, so we have
\begin{eqnarray}   \label{uvexpansion}
 B_\pm(p) &=& \pm
  \frac{32}{3\pi^2 N}\,\frac{\alpha}{p}\, \theta +
 \frac{\alpha^2}{p^2}\left( B_\pm(\alpha) \mp
 \frac{32\theta}{3\pi^2 N} \right) \,,
\end{eqnarray}
for $p > \alpha$.

\subsection{Analytical Results}

So we arrive at the integral equation for momenta $p < \alpha$
\begin{eqnarray}  \label{massintcs}
  B_\pm(p) & = & \frac{32}{3 \pi^2 N} \int_0^\alpha{\rm d}k\,
                \frac{k^2}{k^2 + M_\pm^2}\frac{B_\pm(k)}{\max(p,k)}
    + \frac{8}{3\pi^2 N}
\left(\pm \frac{16\theta}{3\pi^2 N } + B_\pm(\alpha) \right)
\pm \frac{8\theta}{\pi^2 N}  \,,
\end{eqnarray}
which we can now solve analytically by converting it to a second-order
differential equation with boundary conditions.

The second-order differential equation is the same as without the CS
term, Eq.~(\ref{diffmpure}), and also the infrared boundary condition
is the same, the only (but essential!) difference is the
``ultraviolet'' boundary condition at $p = \alpha$
\begin{eqnarray}
   B_\pm(\alpha) + \alpha B'_\pm(\alpha) &=&
\frac{8}{3\pi^2 N}
\left(\pm \frac{16\theta}{3\pi^2 N } + B_\pm(\alpha) \right)
\pm \frac{8\theta}{\pi^2 N} \,,
\end{eqnarray}
which should be compared with Eq.~(\ref{uvbcpure}). It is also
important to keep in mind that the normalization condition is
\begin{eqnarray}
  B_\pm(0) & = & M_\pm \,,
\end{eqnarray}
and that the SD equation does determine the sign of the
mass function as well. This is obviously not the case if $\theta = 0$.

The general solution of the differential equation satisfying the
infrared boundary condition and the normalization condition is
\begin{eqnarray}  \label{gensol}
  B_\pm(p) & = & M_\pm \, _2F_1(a_+,a_-,\frac{3}{2};-p^2/M_\pm^2) \,,
\end{eqnarray}
where $a_\pm = \frac{1}{4}(1 \pm i\sqrt{N_c/N - 1})$.  The
ultraviolet boundary condition leads to the condition
\begin{eqnarray}   \label{uvbccs}
  M_\pm \, _2F_1(a_+,a_-,\frac{1}{2};-\alpha^2/M_\pm^2)
- \frac{8}{3\pi^2 N} M_\pm \, _2F_1(a_+,a_-,\frac{3}{2};-\alpha^2/M_\pm^2)
\nonumber \\  =
\pm \frac{8\theta}{\pi^2 N}\left( 1 + \frac{16}{9\pi^2 N }\right) \,.
\end{eqnarray}

In order to determine $M$, we can plot the LHS of the above equation
divided by $\frac{8}{\pi^2 N}\left( 1 + \frac{16}{9\pi^2 N }\right)$
as function of $M$ for a given value of $N$, which gives us
automatically $\theta$ as function of $M$, see Fig.~\ref{fig6}a. From
this figure we can see that there are three solutions possible for
$B_+$ and $B_-$. Two of them correspond to the two solutions which we
have found numerically, and the third is an oscillating solution
which in the limit of vanishing $\theta$ corresponds to the first
oscillating solution in pure QED, $n=2$ in Eq.~(\ref{pureosc}). For
extremely small values of $M$ and $\theta$ there are more oscillating
solutions, in the region around the origin, but these are numerically
unstable, and correspond to higher excites states, just as in pure
QED.

Using the same notation as for the numerical solution, we have $B_+(0)
> 0$ and $\tilde B_+(0) < 0$, and for $B_-$ the opposite signs. The
set $(B_+, B_-)$ is the chirally symmetric solution which is present
for all values of $(N,\theta)$. The chiral-symmetry-breaking solution
is given by the set $(B_+ , \tilde B_-)$ (or vice versa, the set
$(\tilde B_+ , B_-)$). In the limit $\theta \rightarrow 0$, this
chiral-symmetry-breaking solution behaves in the following way
\begin{eqnarray}
B_e &\rightarrow& m \,,\\
B_o &=& {\cal O}(\theta)\,.
\end{eqnarray}
If we increase $\theta$ from zero at fixed $N$, then the absolute
value of the second solution, $\tilde B_\pm$, decreases (whereas the
other solution increases), until at a critical value $\theta_c$ this
solution coincides with a third solution, and disappears, but does not
become zero at the critical point, see Fig.~\ref{fig6}a. This clearly
signals a first-order phase transition, as was also suggested by the
numerical results. For comparison, we also showed our numerical data
in the same figure, where we use a different value of $\alpha$ for our
numerical and analytical calculation, $\alpha_{\hbox{\scriptsize
num}}$ and $\alpha_{\hbox{\scriptsize an}}$ respectively, in order to
have equal numerical and analytical values of
$m(0)/\alpha_{\hbox{\scriptsize num}}$ in the absence of the CS term
(note that the axes are $\theta/\alpha_{\hbox{\scriptsize num}}$ and
$|B(0)|/\alpha_{\hbox{\scriptsize num}}$ respectively). This shows
clearly that both our numerical and our analytical results are in good
agreement with each other.

We can also plot $M$ versus $N$ for some fixed values of $\theta$,
see Fig.~\ref{fig6}b. From that figure we can see that if we increase
$N$ for fixed $\theta$, the chiral-symmetry-breaking solutions
disappear if $N$ exceeds some critical value $N_c$, which decreases
rapidly as a function of $\theta$. This figure shows that the chiral
phase transition is first order in this direction as well: increasing
$N$ beyond $N_c(\theta)$ makes the second (and third) solution
disappear, but at the phase transition neither $\tilde B$ nor $B_e =
(B_+ + B_-)/2$ (which can be regarded as the order parameter of the
chiral phase transition) becomes zero. In this figure we can also see
that in the limit $\theta \rightarrow 0$ the critical value $N_c$ goes
towards $N_c(0) \simeq 4.32$.

The critical parameters $N_c$ and $\theta_c$ can be calculated by
differentiating the ultraviolet boundary condition,
Eq.~(\ref{uvbccs}), with respect to $M$. This leads to an equation for
$M_c$ as function of $N_c$ (remember that $M$ does not vanish at
$N_c$)
\begin{eqnarray}  \label{mcritcs}
   _2F_1(a_+,a_-,\frac{-1}{2};-\alpha^2/M_c^2)
+ \frac{8}{3\pi^2 N}\;{}_2F_1(a_+,a_-,\frac{1}{2};-\alpha^2/M_c^2)
\nonumber\\
- \frac{16}{3\pi^2 N}\;{}_2F_1(a_+,a_-,\frac{3}{2};-\alpha^2/M_c^2)
 & = & 0 \,,
\end{eqnarray}
which can be used as input for the ultraviolet boundary condition
itself in order to calculate $\theta_c$. Although we do not have an
explicit form for $\theta_c$ as function of $N$, we can calculate it
numerically, and have shown the critical line in Fig.~\ref{fig7}. In
this figure we also show some estimates of the critical parameters
based on our numerical calculation. This shows that the numerical and
analytical results are qualitatively in good agreement with each
other, and the only difference is an overall scale factor (just as in
pure QED).

For small values of $M_c$ and $N_c(\theta)$ close to
$N_c(0) \simeq 4.32$, we can expand the above equation in
$\sqrt{N_c(0)/N-1}$, leading to
\begin{eqnarray}
  M_c/\alpha & = &
    \exp{\left[\frac{-2\,\pi}{\sqrt{N_c(0)/N - 1}}
           +  \frac{1}{2}\pi + 3\,\ln{2} - \frac{8}{21}\right]}
\end{eqnarray}
(see appendix B), which can be inserted into the series expansion for
$\theta$. To leading order in $\sqrt{N_c(0)/N - 1}$, this gives the
following expression for $\theta_c$
\begin{eqnarray}
\theta_c
&\simeq&  \frac{448 \, \Gamma(1/2)}{75 \, \Gamma(1/4)^2 }
   \exp{\left[\frac{-3\,\pi}{\sqrt{N_c/N - 1}}
           +  \frac{3}{4}\pi + \frac{9}{2}\ln{2} - \frac{4}{7}\right]} \,.
\end{eqnarray}

As mentioned before, we can also calculate the chiral condensate which
is well-defined, even without cutoff. The chiral condensate is
\begin{eqnarray}
   \langle\bar\psi\psi\rangle & = &  \frac{1}{\pi^2}
  \int_0^\infty {\rm d}k \left(\frac{k^2\,B_+(k)}{A_+^2(k)k^2 + B_+^2(k)}
  +\frac{k^2\,\tilde B_-(k)}{\tilde A_-^2(k)k^2 + \tilde B_-^2(k)}\right) \,.
\end{eqnarray}
In the approximations we are using here, this reduces to
\begin{eqnarray}
   \langle\bar\psi\psi\rangle & = & \frac{1}{\pi^2}
   \int_0^\alpha {\rm d}k \left(\frac{k^2\,B_+(k)}{k^2 + M_+^2}
+ \frac{k^2\,\tilde B_-(k)}{k^2 + \tilde M_-^2}\right)
+ \frac{1}{\pi^2} \int_\alpha^\infty {\rm d}k  (B_+(k)+ \tilde B_-(k))
\end{eqnarray}
which can be calculated analytically, using Eq.~(\ref{massintcs}) at
$p = \alpha$ and the ultraviolet expansion Eq.~\ref{uvexpansion} for
$B_\pm$. This leads to
\begin{eqnarray}
   \langle\bar\psi\psi\rangle & = & \frac{\alpha}{\pi^2}
 \Big(B_+(\alpha) + \tilde B_-(\alpha)\Big)
  \left(\frac{3 \pi^2 N}{32} + \frac{3}{4}\right)
\end{eqnarray}
with $B_+(\alpha)$ and $\tilde B_-(\alpha)$ determined through
Eqs.~(\ref{gensol}) and (\ref{uvbccs}). In Fig.~\ref{fig8} we show the
chiral condensate (together with the $B_e(0)$) as obtained both
numerically and analytically for $\theta = 0$ and $\theta = 10^{-5}$
as a function of $N$. Again, they are qualitatively in good agreement
with each other, and show a discontinuity in both the chiral
condensate and $B_e(0)$ at the critical point.

In Fig.~\ref{fig9}, we show both $B_e(p)$ and $B_o(p)$, as obtained
numerically and analytically. The numerical and the analytical
solutions have the same behavior in both the infrared and the
ultraviolet region; around $p = \alpha$ there is of course a kink in
the analytical solution due to the approximations we have made.
There is a scale difference between the analytical and the numerical
solution for the chiral-symmetry-breaking solution, just as in pure
QED; in fact, if we compare this figure with Fig.~\ref{fig2}, we can
see that the parity-even mass function is almost the same as the
dynamical mass function in pure QED.  Furthermore, this figure shows
very well the difference in the ultraviolet between the even and the
odd mass function: the odd mass function behaves like $\theta/p$,
whereas the even mass function $B_e(p) = (B_+(p) + \tilde B_-(p))/2$
behaves like $1/p^2$, due to the cancellation of the terms
proportional to $\pm\theta/p$.

\section{CJT Effective Potential}
\label{CJTefpot}

We consider the effective action of Cornwall--Jackiw--Tomboulis
\cite{CJT74} (CJT) which is given by
\begin{eqnarray}
 \Gamma[S] &=& \Gamma_0[S] + \Gamma_1[S]\,,
\end{eqnarray}
with
\begin{eqnarray}
 \Gamma_0[S] &\equiv&
- {\rm Tr}\big[{\rm Ln} [S^{-1} S_0] + S_0^{-1}S  - 1\big]\,,
\\
 \Gamma_1[S] &\equiv& {e^2 \over 2}
    {\rm Tr}[S\gamma_\mu D_{\mu\nu} \gamma_\nu S]\,,
\end{eqnarray}
where $D_{\mu\nu}$ is the full photon propagator, and $S_0$ and $S$
are the bare and the full fermion propagator respectively. At the
stationary point $\delta \Gamma[S]/\delta S=0$, this effective action
actually gives the SD equation: $S^{-1}-S_0^{-1} = -e^2 {\rm Tr}
[\gamma_\mu D_{\mu\nu}\gamma_\nu S]$, as is shown explicitly in
appendix C. Note that we have chosen the bare vertex to write the
effective action, which can be justified by taking the nonlocal gauge.
Furthermore we remark that the effective action is normalized so that
$\Gamma_0[S=S_0] = 0.$ For a review of the effective potential, see
e.g. \cite{Haymaker91}.

We can rewrite the above expression in terms of the
fermion propagators $S_{+}$ and $S_{-}$
\begin{equation}
 S = S_{+} \chi_{+} + S_{-} \chi_{-}\,.
\end{equation}
By using the fact that $\chi_{+}$ and $\chi_{-}$ are projection operators
and the relation $\hbox{Tr Ln} = \hbox{Ln Det}$, it is not difficult
to show that both $\Gamma_0[S]$ and $\Gamma_1[S]$ are decomposed into
two parts
\begin{eqnarray}
  \Gamma_{0,1}[S] &=& \Gamma_{0,1}[S_{+}]+\Gamma_{0,1}[S_{-}]\,,
\end{eqnarray}
where
\begin{eqnarray}
  \Gamma_0[S_{\pm}] &\equiv& - {\rm Tr}\big[{\rm Ln} [S_{\pm}^{-1} S_0]
  + S_0^{-1}S_{\pm}  - 1 \big]\,,
\end{eqnarray}
and
\begin{eqnarray}
\Gamma_1[S_{\pm}] &\equiv& {e^2 \over 2}
 \int {{{\rm d}^3 p} \over (2\pi)^3}  \int {{{\rm d}^3 k} \over (2\pi)^3}
{\rm Tr}[\gamma^\mu S_{\pm}(p) \gamma^\nu S_{\pm}(k)
\chi_{\pm}] D_{\mu\nu}(k-p) \,.
\end{eqnarray}

Now we define the effective potential $V[S]$ by dividing the effective
action by the space-time volume $\int d^3x$: $V[S] \equiv
\Gamma[S]/\int d^3x$. In momentum space, $V_0[S_{\pm}]$ is given by
\begin{eqnarray}
 V_0[S_{\pm}] &=& {N \over 2\pi^2} \int_0^\infty p^2 {\rm d}p
 \Biggr[ - \ln \left( {p^2A_{\pm}^2(p)+B_{\pm}^2(p) \over
p^2}\right)
\nonumber\\&&
 + 2 {p^2 A_{\pm}(p)[A_{\pm}(p)-1]+B_{\pm}^2(p)
\over p^2A_{\pm}^2(p)+B_{\pm}^2(p)}
\Biggr] \,.
\end{eqnarray}
For $V_1[S_{\pm}]$, after calculating the trace in the integrand, we
arrive at the result
\begin{eqnarray}
{\lefteqn{ V_1[S_{\pm}]
=  e^2 \int {{\rm d}^3p \over (2\pi)^3}\int {{\rm d}^3k \over (2\pi)^3}
{1 \over [p^2A_{\pm}^2(p)+B_{\pm}^2(p)][k^2A_{\pm}^2(k)+B_{\pm}^2(k)]} }}
\nonumber\\&&
\times \Biggr\{
\left(2\,D_T(q) {(q \cdot p) \, (q \cdot k) \over q^2}
+ \frac{a(q)}{q^4}
{\Big(2 (q \cdot p) \, (q \cdot k) - p\cdot k\Big)}\right) A_{\pm}(p)A_{\pm}(k)
\nonumber\\&&
- \Big(2\,D_T(q) + \frac{a(q)}{q^2}\Big) B_{\pm}(p)B_{\pm}(k)
\pm {D_O(q) \over |q|} [q \cdot p A_{\pm}(p)
B_{\pm}(k) - q \cdot k B_{\pm}(p) A_{\pm}(k)] \Biggr\}\,,
\end{eqnarray}
with $q = k - p$.
At this stage we adopt the nonlocal gauge $a(q)$ which leads to no
wave-function renormalization for the fermion: $A_{\pm}(p) \equiv 1$.
Then we can write
\begin{equation}
 V[B] = V[S]\Big|_{A_{\pm}\equiv 1} =
V_0[B_{+}]+V_0[B_{-}]+V_1[B_{+}]+V_1[B_{-}]\,,
\end{equation}
where
\begin{equation}
 V_0[B_{\pm}] = {N \over 2\pi^2} \int_0^\infty p^2 {\rm d}p
 \left[ - \ln \left( {p^2+B_{\pm}^2(p) \over
p^2}\right) + 2 {B_{\pm}^2(p)
\over p^2+B_{\pm}^2(p)}
\right] \,,
\end{equation}
and
\begin{eqnarray}
\lefteqn{V_1[B_{\pm}]
=  e^2 \int {{\rm d}^3p \over (2\pi)^3}\int
{{\rm d}^3k \over (2\pi)^3} {1 \over
[p^2+B_{\pm}^2(p)][k^2+B_{\pm}^2(k)]}}
\nonumber\\&&
\times \Biggr\{
\left(-2\,D_T(q)  - \frac{a(q)}{q^2}\right) B_{\pm}(p)B_{\pm}(k)
\pm {D_O(q) \over |q|} [(q \cdot p) B_{\pm}(k)
- (q \cdot k) B_{\pm}(p) ] \Biggr\}\,.
\end{eqnarray}
Actually this reproduces the SD equation for $B_{\pm}$ in the
nonlocal gauge at the stationary point $\delta V[B]/\delta B_{\pm}=0$
as shown in appendix C.

At the stationary point of $V[B]$, $B_{\pm}$ satisfies the SD equation
and hence $B_{\pm}$ is given by the solution to the SD equation:
$B_{\pm}=B_{\pm}^{sol}$. Therefore, at the stationary point, we obtain
\begin{eqnarray}
V_1[B_{\pm}]
 &=&  -  \int {{\rm d}^3p \over (2\pi)^3}
 {B_{\pm}(p)[B_{\pm}(p)-m_{\pm}] \over  p^2+B_{\pm}^2(p)} \,.
\end{eqnarray}
Hence the effective potential at the stationary point is given by
\begin{eqnarray}
V[B_{sol}] &=&
V[B_{+}^{sol}]+V[B_{-}^{sol}]\,,
\nonumber\\
V[B_{\pm}]
&=& -\int {{\rm d}^3p \over (2\pi)^3} \left[ \ln \left( 1 +
{B_{\pm}^2(p) \over p^2} \right)
 - {B_{\pm}(p)[B_{\pm}(p)+m_{\pm}] \over  p^2+B_{\pm}^2(p)} \right] \,.
\end{eqnarray}
Note that the function $g(x)=\ln(1+x)-{x \over 1+x}$ is positive and
monotonically increasing in $x(>0)$.  Therefore, when $m_{\pm}=0$,
\begin{equation}
V[B_{sol}]=- {1 \over 2\pi^2} \int_0^\infty p^2 {\rm d}p
\left[  g\left({B_{+}^2(p) \over p^2}\right)
+g\left({B_{-}^2(p) \over p^2}\right) \right]
\label{EPfinal}
\end{equation}
is non-positive, $V[B_{sol}] \le 0$, and all the nontrivial solutions
have lower energy than the trivial ones (in the case of pure QED).
Therefore we can determine the ground state, namely the solution which
minimizes the effective potential, by using Eq.~(\ref{EPfinal}).

Without actually calculating the integral, we can easily conclude that
in the presence of an explicit CS term, the chirally symmetric
solution $(B_{+}$,$B_{-})$ gives a lower effective potential and is
thus favored above the chiral-symmetry-breaking solution
$(B_{+}$,$\tilde B_{-})$. Since the effective potential is the sum of
two non-positive integrals, each of which depends on $B_+^2$ or
$B_-^2$, this effective potential is minimized by a symmetric set of
solutions $(B_{+}$,$B_{-})$ and not by $(B_{+}$,$\tilde B_{-})$.  So
in the presence of an explicit CS term, the chirally symmetric phase
is the ground state, even if the number of fermion flavors $N$ and the
CS parameter $\theta$ are below their critical values for dynamical
chiral-symmetry breaking. This is quite surprising, since it is
well-known that without the explicit CS term the chiral symmetry is
broken for $N$ below the critical number, as can also be seen from the
effective potential, Eq.~(\ref{EPfinal}).

In solving the SD equation we have put the external source $m_e$ and
$m_o$ to zero from the beginning.  As a result, the solution for the
homogeneous SD equation has no specific direction for the dynamical
mass to be generated. Actually, if a solution is found for $B_+$, then
another solution $B_-=-B_+$ is automatically obtained, and it is this
solution which gives the lowest effective potential. In order to study
the {\em spontaneous} breaking of a symmetry in general, one can
introduce an external source $J{\cal O}$ which breaks the symmetry in
question and subsequently consider the limit of removing the external
source.  If the symmetry is broken even in this limit, which is
signaled by the non-vanishing order parameter $\phi = \lim_{J
\rightarrow 0}\langle {\cal O} \rangle_J $, then it is said that the
symmetry is spontaneously broken.  In taking this limit one must
specify from which direction the external source is decreased to zero,
$J \rightarrow 0$.

In this case we must consider the two limits: $m_e \rightarrow 0$,
$m_o \rightarrow 0$.  For simplicity we keep $m_e$ and $m_o$ positive,
$m_e, m_o \ge 0$ and consider the limit: $m_e \downarrow 0$, $m_o
\downarrow 0$, without loss of generality.  Then there are five cases
to be examined, which are given below.
\bigskip

\begin{tabular}{l@{\extracolsep{6\tabcolsep}}lll}
\multicolumn{2}{c}{explicit masses} &
\multicolumn{1}{c}{condensates} &
\multicolumn{1}{c}{solutions of SD eq.}\\ \hline \vspace{-\bigskipamount}\\
$m_e > m_o > 0$   &
$m_+ > m_- > 0$   &
$\langle\bar\psi\psi\rangle_+ > \langle\bar\psi\psi\rangle_- > 0$ &
$B_+(0) > 0$, $\tilde B_-(0) > 0$                                 \\

$m_e > 0$, $m_o = 0$  &
$m_+ = m_- > 0$       &
$\langle\bar\psi\psi\rangle_+ = \langle\bar\psi\psi\rangle_- > 0$ &
$B_+(0) > 0$, $\tilde B_-(0) > 0$                                 \\

$m_o > m_e > 0 $     &
$m_+ > -m_- > 0 $    &
$\langle\bar\psi\psi\rangle_+ > 0 > \langle\bar\psi\psi\rangle_- $ &
$B_+(0) > 0$, $B_-(0) < 0$                                        \\

$m_o > 0$, $m_e = 0 $    &
$m_+ = -m_- > 0 $        &
$\langle\bar\psi\psi\rangle_+ = - \langle\bar\psi\psi\rangle_- > 0$ &
$B_+(0) = - B_-(0) >0$                              \\

$m_o = m_e > 0$            &
$m_+ > 0$, $m_- = 0 $      &
$\langle\bar\psi\psi\rangle_+ > 0$, &
$B_+(0) > 0$,      \\
       &
       &
$\langle\bar\psi\psi\rangle_-$  undetermined &
both $B_-(0)$ and $\tilde B_-(0)$
\end{tabular}
\bigskip\\
Therefore the two solutions $(B_+,B_-)$ and $(B_+,\tilde B_-)$ are
realized in two different limits: the limit $m_+ = -m_-
\downarrow 0$ leads to the solution $(B_+,B_-)$, whereas the limit
$m_+ = m_- \downarrow 0$ yields the solution $(B_+,\tilde B_-)$. This
last set of solutions is of course only possible if the parameters
$N,\theta$ are in the chirally broken phase.

{}From this consideration, the two sets of solutions cannot be realized
simultaneously. The result which solution is realized depends on the
ordering of taking the vanishing limit keeping the relation between
$m_e$ and $m_o$. So in the presence of the CS term, taking the limit
$m_e > m_o \downarrow 0$, we will get the solution $(B_+,\tilde B_-)$
which has a higher value for the CJT effective potential than that of
the ``trivial'' solution $(B_+,B_-)$. That means that this solution is
a quasi-stable solution of the SD equation, and will eventually decay
into the ``trivial'' solution. This quasi-stable solution does no
longer exist beyond a certain value of $\theta$, the critical value
$\theta_c(N)$ depending on $N$.  At this point the system jumps
directly to the state expressed by the solution $(B_+,B_-)$, which is
energetically more favored, in a discontinuous way.

\section{Conclusion and Discussion}

In (2+1)-dimensional QED with $N$ flavors of four-component Dirac
fermions, we have solved the SD equation for the fermion propagator in the
nonlocal gauge both analytically and numerically.  In the absence of the
bare CS term, we have shown the existence of a finite critical number of
flavors $N_c \cong 4.3$, below which the chiral symmetry is spontaneously
broken, in agreement with previous analyses
\cite{ANW88,Nash89,KEIT94}.  In the presence of CS term, $\theta\not=0$,
we have obtained the critical line extending from the critical point
$(N,\theta)=(N_c,0)$ in the two-dimensional phase diagram $(N,\theta)$.
Here a quite remarkable point is that, no matter how small $\theta$ is,
this phase transition turns into a first-order transition when
$\theta\not=0$, although the critical point $(N_c,0)$ in the absence of CS
term is a continuous (infinite-order) phase transition point. Therefore the
point $(N_c,0)$ is the only one point on the critical line which exhibits
continuous phase transition.

We have shown that the chiral-symmetric solution $(B_+, B_-)$ and the
chiral-symmetry-breaking one $(B_+, \tilde B_-)$ of the SD equation are
stationary points of the effective potential of the CJT type. Here it
should be remarked that the trivial solution $(B_+ \equiv 0, B_-\equiv 0)$
as a trivial stationary point (equivalently the trivial solution of the SD
equation) is possible only when $\theta=0$. Although the symmetric
solution $(B_+, B_-)$ gives lower effective potential than the
symmetry-breaking solution $(B_+,\tilde B_-)$, we have shown that it is
possible to take the limit of removing the external source, $m_e, m_o
\downarrow 0$, so that the {\em spontaneously} chiral-symmetry-breaking
solution $(B_+, \tilde B_-)$ is realized.  However it is not yet clear
whether they give local minima, local maxima, or possibly a saddle
point of the effective potential and whether or not $(B_+, B_-)$ gives
the absolute minimum. In order to confirm this issue, it is necessary
to calculate the second functional derivative:
\begin{equation}
 {\delta^2  \over \delta B_\pm(p) \delta B_\pm(q)} V_{CJT}[B_+,B_-]
\end{equation}
at the respective stationary point \cite{Matsuki91}. The system at an
excited state
$(B_+, \tilde B_-)$ might be quasi-stable and might decay into the more
stable state $(B_+, B_-)$ in a finite time interval. The stability of the
stationary point will be discussed in more detail in a subsequent paper.

It is interesting to see our result from the viewpoint of the lattice
gauge theory where the lattice spacing $a$ corresponds to the inverse of
the UV cutoff $\Lambda$, $a \sim 1/\Lambda$.  It is well known
\cite{Creutz83} that the nontrivial continuum limit can be possibly taken
only at the second- and higher-order transition point where the
correlation length diverges (in units of the lattice spacing).  Hence our
result implies that the meaningful {\it continuum} limit from the broken
phase in the nonperturbatively regularized three-dimensional gauge field
theory can be taken only at the point $(N_c,0)$, i.e. in the absence of the
bare CS term.  This is a novel feature of the three-dimensional gauge
field theory with a bare CS term, which seems to be overlooked so far. Our
result may have important implications in the application of the
three-dimensional gauge theory in condensed-matter physics as a
long-wavelength effective theory of the microscopic model. This will be
discussed in a forthcoming paper.

This result is obtained by using a consistent expansion of the full
gauge boson propagator and the vertex in $1/N$. In order to satisfy
the WT identity (up to terms of order $1/N$ and of order of the
dynamically generated mass), we have adopted the nonlocal gauge
function which allows us to neglect effect from the wave-function
renormalization.  A more accurate approximation would include a full
$1/N^2$ calculation, similar to the analysis in \cite{Nash89} in pure
QED. Since in pure QED these $1/N^2$ corrections do not change the
result qualitatively, but only quantitatively (leading to a slightly
different critical number of fermion flavors), we do not expect that
those corrections will change our present result essentially. The
question of gauge covariance can (in principle) be recovered by
applying the Landau--Khalatnikov transformation rules to the various
Green's functions
\cite{lk55}. Therefore we expect our present results to hold also in a
more elaborate approximation of the SD equation.

\section*{Acknowledgements}

This work was initiated during a visit of one of the authors (P.~M.)
to Chiba University, and he would like to thank the members of the
Graduate School of Science and Technology for their hospitality during
that stay.  We would like to thank K.~Yamawaki, Yoonbai Kim, and
T.~Ebihara for stimulating discussions, and A.~Shibata for helping
with printing the postscript figures. P.~M. has been financially
supported by the JSPS (being a JSPS fellow under number 94146) and
K.-I.~K. is partly supported by the Computation Centre of Nagoya
University.

\appendix

\section{Integrations}

\subsection{Angular Integration Formulae}
First we note the following relation
\begin{equation}
\int_0^\pi {\rm d} \phi \sin \phi f(q) = \int_{-1}^{1} {\rm d}z f(q)
= {1 \over pk} \int_{|p-k|}^{p+k} q {\rm d}q f(q) \,,
\end{equation}
where $q^2 = (p-k)^2 = p^2+k^2-2pk z$ with $z = \cos \phi$.
Then we obtain the following formulae
\begin{eqnarray}
\int_{-1}^{1} {\rm d}z {1 \over q} &=& {2\min{(p,k)} \over pk} \,,
\\
\int_{-1}^{1} {\rm d}z {1 \over q^2}
&=& {1 \over pk} \ln {p+k \over |p-k|} \,,
\\
\int_{-1}^{1} {\rm d}z {1 \over q^3}
&=& {1 \over pk} \left( {1 \over |p-k|}  - {1 \over p+k}
\right) \,,
\\
\int_{-1}^{1} {\rm d}z {1 \over q^4}
\ln \left(1+{q \over \alpha}\right)
&=& {1 \over pk} \Biggr[
{1 \over 2\alpha} \left({1 \over |p-k|}-{1 \over p+k}\right)
- {1 \over 2\alpha^2} \ln {1+{\alpha \over
|p-k|} \over
1+{\alpha \over p+k}}
\nonumber\\&&
+ {1 \over 2|p-k|^2}\ln \left(1+{|p-k| \over \alpha}\right)
- {1 \over 2(p+k)^2}\ln \left(1+{p+k \over \alpha}\right)
\Biggr] \,,
\\
\int_{-1}^{1} {\rm d}z {1 \over q(q+\alpha)}
&=& {1 \over pk} \ln {p+k+\alpha \over |p-k|+\alpha} \,.
\end{eqnarray}
For the calculation of the explicit CS term it is more convenient to use
\begin{eqnarray}
\int_{-1}^{1} {\rm d}z {1 \over {q^2 + p^2 + 2pqz + M^2}}
&=& {1 \over {2pq}} \ln {{(p+q)^2 + M^2} \over {(p-q)^2 + M^2}} \,,
\\
\int_{-1}^{1} {\rm d}z {{p\,q\,z} \over {q^2 + p^2 + 2pqz + M^2}}
&=& 1 + {{p^2 + q^2 + M^2} \over {4\,p\,q}}
 \ln {{(p+q)^2 + M^2} \over {(p-q)^2 + M^2}} \,.
\end{eqnarray}
In pure QED this gives for the angular integration
\begin{eqnarray}
 {\rm K}(p,k) &=&  \int_{-1}^{1} {\rm d}z
\left({2\over {{q}\, \left( \alpha  + {q} \right) }}
       +  {{2\,\alpha }\over {q^3}}  - {{1}\over {q^2}}
  - {{2\,{{\alpha }^2}\,
\ln \left({{\alpha  + {q}}\over {\alpha }}\right)}\over  {q^4}} \right)
\\
&=&
{{2\,\alpha }\over {\max(p,k)\,{|k^2 - p^2 |}}}
+ \frac{1}{k\,p} \ln{\frac{\alpha + | k + p|}{\alpha + | k - p |}}
\nonumber\\ &&
- {{{\alpha^2}\,\ln (1 + {{{| k - p |}}\over \alpha})}\over
{k\,p\,{{\left( -k + p \right) }^2}}}
+ {{{\alpha^2}\,\ln (1 + {{{| k + p |}}\over \alpha})}\over
{k\,p\,{{\left( k + p \right) }^2}}} \,.
\end{eqnarray}

\subsection{Explicit Chern--Simons Term}

The explicit CS term, proportional to
\begin{eqnarray}
 I(p)&\equiv& \int_0^\infty \frac{k^2 \, {\rm
d}k}{k^2+M_\pm^2}
       \int_{-1}^1{\rm d}z \frac{k \cdot (k-p)}{(k-p)^2(|k-p| + \alpha)^2}
\nonumber\\
&=&   \int_0^\infty \frac{{\rm d}q}{(q + \alpha)^2}
       \int_{-1}^1{\rm d}z \frac{(q+p) \cdot q}{(q+p)^2+M_\pm^2} \,,
\end{eqnarray}
can be calculated analytically. The angular integration gives
\begin{eqnarray}
 I(p) &=& \int_0^\infty \frac{{\rm d}q}{(q +
\alpha)^2}
       \left(1 - \frac{p^2 - q^2 + M_\pm^2}{4\,p\,q}
     \ln{\frac{(p+q)^2 + M_\pm^2}{(p-q)^2 + M_\pm^2}} \right) \,,
\end{eqnarray}
which we can calculate approximately by expanding the logarithm for $p
< q$ and $p > q$. Taking into account only the leading-order terms in
$\min(p,q)/\max(p,q)$ gives
\begin{eqnarray}
 I(p) &=& \frac{3\,M_\pm^2 + p^2}{3(M_\pm^2 + p^2)^2}
       \int_0^p \frac{2\,q^2\,{\rm d}q}{(q + \alpha)^2}
+   \int_p^\infty \frac{{\rm d}q}{(q + \alpha)^2}
       \frac{2 q^2}{M_\pm^2 + q^2} \,,
\end{eqnarray}
which can easily be calculated. Expanding the result for $M_\pm \ll
\alpha$ gives
\begin{eqnarray}
 I(p) &=& 2\, \frac{3\,M_\pm^2 + p^2}{3(M_\pm^2 + p^2)^2}
 \left(\frac{p^2 + 2\alpha p}{\alpha+p} - 2\alpha\ln{(1+p/\alpha)}\right)
\nonumber \\ &&
 + 2 \Bigg( {{1}\over {\left( \alpha + p \right) }}
 - {{M_\pm}\over {\alpha^2}}
   \left( {{\pi}\over {2}} - \arctan \left({p\over M_\pm}\right)\right)
 + \frac{M_\pm^2}{\alpha^3}\,\ln{\frac{(\alpha + p)^2}{({M_\pm^2} + {p^2})}}
 \Bigg) \,.
\end{eqnarray}
However, we are only interested in the leading-order behavior in the
infrared and ultraviolet region, so we finally get for $p \ll \alpha $
\begin{eqnarray}
 I(p) &=& 2 + {\cal O}(p/\alpha) + {\cal O}(M_\pm/\alpha) \,,
\end{eqnarray}
or $p \gg \alpha \gg M_\pm$
\begin{eqnarray}
 I(p) &=& \frac{8 \alpha}{3 p}
       + {\cal O}(\alpha^2/p^2) + {\cal O}(M_\pm/ p) \,.
\end{eqnarray}


\section{UV Expansion of the Hypergeometric Function}

\subsection{Pure QED}

In pure QED, the ultraviolet boundary condition, Eq.~\ref{uvbcpure},
leads to the requirement
\begin{eqnarray}  \label{uvbcgen}
 _2F_1(a_+,a_-,\frac{3}{2};\frac{-\alpha^2}{m^2})
  + \alpha \,\,_2F_1'(a_+,a_-,\frac{3}{2};\frac{-\alpha^2}{m^2})
 &=&
\frac{8}{3 \pi^2 N}\, _2F_1(a_+,a_-,\frac{3}{2};\frac{-\alpha^2}{m^2})\,,
\end{eqnarray}
with $a_\pm = \frac{1}{4} \pm \frac{1}{4}i\sqrt{N_c/N - 1}$ and
$N_c = 128/(3\pi^2)$. The derivative w.r.t. $\alpha$ of the
hypergeometric function gives
\begin{eqnarray}
 _2F_1(a_+,a_-,\frac{3}{2};\frac{-\alpha^2}{m^2})
  + \alpha \,\,_2F_1'(a_+,a_-,\frac{3}{2};\frac{-\alpha^2}{m^2})
 &=&
 _2F_1(a_+,a_-,\frac{1}{2};\frac{-\alpha^2}{m^2})  \,,
\end{eqnarray}
so we arrive at
\begin{eqnarray}
  _2F_1(a_+,a_-,\frac{1}{2};\frac{-\alpha^2}{m^2})
 &=& a_+ a_- \,
  _2F_1(a_+,a_-,\frac{3}{2};\frac{-\alpha^2}{m^2}) \,,
\end{eqnarray}
where we have used $8/(3\pi^2 N) = a_+ a_-$.
To derive the behavior of the mass $m$ close to the critical value, we
expand Eq.~(\ref{uvbcgen}) in $\alpha/m$. Taking into account leading
order only, we find
\begin{eqnarray}
 _2F_1(a_+,a_-,\frac{1}{2};\frac{-\alpha^2}{m^2}) &=&
\frac{\Gamma(1/2)\Gamma(a_- - a_+)}{\Gamma(a_-)\Gamma(1/2 - a_+)}
\left(\frac{\alpha}{m}\right)^{-2a_+} + {\rm h.c.} \,\,,\\
 _2F_1(a_+,a_-,\frac{3}{2};\frac{-\alpha^2}{m^2})  &=&
\frac{\Gamma(3/2)\Gamma(a_- - a_+)}{\Gamma(a_-)\Gamma(3/2 - a_+)}
\left(\frac{\alpha}{m}\right)^{-2a_+} + {\rm h.c.}  \,\,.
\end{eqnarray}
We are interested in $m/\alpha$, so we rewrite the boundary condition
into
\begin{eqnarray}   \label{moverapure}
\left(\frac{m}{\alpha}\right)^{i\omega}
    &=&
\frac{\Gamma(1 + \frac{i}{2} \omega)\Gamma(a_-)^2 (1 - \frac{1}{2}a_-)}
     {\Gamma(1 - \frac{i}{2} \omega)\Gamma(a_+)^2 (1 - \frac{1}{2}a_+)}
         \,,
\end{eqnarray}
using the notation $\omega = \sqrt{N_c/N - 1}$.  Since the absolute
value of the RHS is equal to one, we can write it as
$e^{2 i (\phi - n \pi)}$, with
\begin{eqnarray}
 \phi & = & \arg{\left[{\Gamma(1 + \textstyle{\frac{i}{2}} \omega)}
       {\Gamma(a_-)^2}\left({1 - \textstyle{\frac{1}{2}}a_-}\right)\right]}
   \,,
\end{eqnarray}
and $n$ integer.  This leads to the equation for $m/\alpha$
\begin{eqnarray}  \label{eqappm}
\frac{m}{\alpha}&=& \exp{\frac{-2n\pi+ 2\phi}{\omega} } \,.
\end{eqnarray}
Next, we expand $\phi$ leading order in $\omega$, using a Taylor
expansion and the Euler constant $\gamma$
\begin{eqnarray}
\Gamma(1 \pm \textstyle{\frac{i}{2}} \omega)
 &=& 1 \mp \textstyle{\frac{i}{2}} \omega \gamma
         + {\cal O}(\omega^2)
\\
\Gamma(a_\pm)
 &=& \Gamma(1/4) \bigg( 1 \mp ( \gamma + \textstyle{\frac{1}{2}}\pi
   + 3\ln{2}) \textstyle{\frac{i}{4}}\omega \bigg)
          + {\cal O}(\omega^2) \,,
\end{eqnarray}
which leads to
\begin{eqnarray}
 \phi & \simeq & \left( \frac{1}{4}\pi + \frac{3}{2}\ln{2}
              + \frac{1}{7} \right) \omega  \,.
\end{eqnarray}
In this way we arrive finally at
\begin{eqnarray}  \label{eqappmexp}
 \frac{m}{\alpha}  &\simeq& \exp{\left[\frac{-2n\pi}{\omega}
         +  \frac{1}{2}\pi + 3\,\ln{2} + \frac{2}{7}\right]} \,.
\end{eqnarray}
This expression is only valid for $m \ll \alpha$, which means that $n$
has to be positive. Using the effective potential, one can show that
the largest value for $m/\alpha$ gives the lowest energy, therefore
the ground state corresponds to $n=1$, and higher excited
(oscillating) solutions are given by $n = 2, 3, 4, \ldots$.

For the chiral condensate we need to know the mass function at
momentum $p = \alpha$, $m(\alpha)$. Using the same expansions as
above, we have
\begin{eqnarray}
  m(\alpha) & = & m \left(
\frac{2i\Gamma(3/2)\Gamma(1-\frac{i}{2}\omega)}
     {\omega\Gamma(a_-)\Gamma(3/2 - a_+)}
\left(\frac{\alpha}{m}\right)^{-2a_+} + {\rm h.c.} \right) \,,
\end{eqnarray}
with $m$ given by Eq.~(\ref{eqappm}). Close to the critical coupling we
can expand this in $\omega$, using
\begin{eqnarray}
\left(\frac{m}{\alpha}\right)^{\frac{i}{2}\omega}
 & = & e^{i(-\pi+\phi)}
 \;=\; -(\cos{\phi} + i\sin{\phi}) \,,
\end{eqnarray}
which leads to
\begin{eqnarray}
  \frac{m(\alpha)}{\alpha} & \simeq &
 - 8 \,\frac{ \Gamma{(1/2)}}{ \Gamma{(1/4)}^2}
 \left(\frac{m}{\alpha} \right)^{3/2}
 \left( \left(- 1 + \frac{\pi}{4} + \frac{3}{2}\ln{2}\right) \cos{\phi}
  - \frac{\sin{\phi}}{\omega} \right) \,.
\end{eqnarray}
Finally, using the expansion for $\phi$, we find
\begin{eqnarray}
  \frac{m(\alpha)}{\alpha} & \simeq &
  \frac{64}{7}
  \frac{ \Gamma{(1/2)}}{ \Gamma{(1/4)^2}}
  \left(\frac{m}{\alpha} \right)^{3/2}  \,,
\end{eqnarray}
with $m/\alpha$ given by Eq.~(\ref{eqappmexp}).

\subsection{Chern--Simons Term: critical mass as function of $N$}

The equation for $M_c$ as function of $N$, Eq.~\ref{mcritcs}, is
\begin{eqnarray}
   _2F_1(a_+,a_-,\frac{-1}{2};-\alpha^2/M_c^2)
 + a_+a_- \;{}_2F_1(a_+,a_-,\frac{1}{2};-\alpha^2/M_c^2)
\nonumber\\
 - 2a_+a_- \;{}_2F_1(a_+,a_-,\frac{3}{2};-\alpha^2/M_c^2)
 & = & 0 \,.
\end{eqnarray}
We can expand this as usual for small $M_c$
\begin{eqnarray}
\left(\frac{\alpha}{M_c}\right)^{-2a_+}
\Bigg(\frac{\Gamma(-1/2)\Gamma(a_- - a_+)}{\Gamma(a_-)\Gamma(-1/2 - a_+)}
+ a_+a_-
\frac{\Gamma(1/2)\Gamma(a_- - a_+)}{\Gamma(a_-)\Gamma(1/2 - a_+)}
\nonumber\\
- 2a_+a_-
  \frac{\Gamma(3/2)\Gamma(a_- - a_+)}{\Gamma(a_-)\Gamma(3/2 - a_+)}
\Bigg) + {\rm h.c.}
  &=&  0 \,,
\end{eqnarray}
which can be reduced to
\begin{eqnarray}
\left(\frac{\alpha}{M_c}\right)^{-2a_+}
\frac{\Gamma(a_- - a_+)}{\Gamma(a_-)^2}
\left( 1 + a_+ + a_+a_- \right) + {\rm h.c.}
  & = & 0 \,,
\end{eqnarray}
so we arrive at
\begin{eqnarray}
\left(\frac{M_c}{\alpha}\right)^{i\omega}
& = &
\frac{\Gamma(1 + \textstyle{\frac{i}{2}}\omega)
               \Gamma(a_-)^2 (1 + a_- + a_+a_-)}
     {\Gamma(1 - \textstyle{\frac{i}{2}}\omega)
               \Gamma(a_+)^2 (1 + a_+ + a_+a_-)}  \,.
\end{eqnarray}
Again, we can write this as
\begin{eqnarray}
\frac{M_c}{\alpha}&=& \exp{\frac{-2\,n\,\pi+ 2\phi}{\omega} } \,,
\end{eqnarray}
but now with
\begin{eqnarray}
\phi & = &
\arg[{\Gamma(1 + \textstyle{\frac{i}{2}}\omega)
               \Gamma(a_-)^2 (1 + a_- + a_+a_-)}] \,.
\end{eqnarray}
Finally we expand $\phi$ in $\omega$
\begin{eqnarray}  \label{eqappphics}
\phi  & \simeq &
\left( \frac{1}{4}\pi + \frac{3}{2}\ln{2}
    - \frac{4}{21} \right)\omega  \,,
\end{eqnarray}
so we arrive at
\begin{eqnarray}
\frac{M_c}{\alpha}  &=&
  \exp{\left[\frac{-2\,n\,\pi}{\omega}
           +  \frac{1}{2}\pi + 3\,\ln{2} - \frac{8}{21}\right]} \,,
\end{eqnarray}
with $n = 1,2,3,..$; the term $-2\,n\,\pi/\omega$ arises
just as in pure QED, and also here there are infinitely many solutions
for each value of $N$. The interesting one is the one corresponding to
the ground state in pure QED, $n = 1$.

\subsection{Chern--Simons Term: critical $\theta$ as function of $N$}

Using the above approximation for $M_c$ as function of $N$, we can now
calculate $\theta_c$ as function of $N$, see Eq.~\ref{uvbccs}. In
general we have
\begin{eqnarray}
\theta_c &=&
 \frac{\pm \pi^2 N}{8(1 + \frac{16}{9\pi^2 N})}
 \Bigg( M_c \, _2F_1(a_+,a_-,\frac{1}{2};-\alpha^2/M^2_c)
 - a_+a_- M_c \, _2F_1(a_+,a_-,\frac{3}{2};-\alpha^2/M^2_c)
 \Bigg) \,,
\end{eqnarray}
which we can expand
\begin{eqnarray}
\frac{\theta_c}{\alpha} &=&  \pm
\frac{ \pi^2 N\,\Gamma{(1/2)}}{8  (1 + \frac{16}{9\pi^2 N })}
  \left(\frac{M_c}{\alpha}\right)^{3/2}
\left(\left(\frac{M_c}{\alpha}\right)^{\frac{i}{2}\omega}
\frac{\Gamma{(a_- - a_+)}}{\Gamma{(a_-)}^2}
 (1-\textstyle{\frac{1}{2}}\,a_+)
 + {\rm h.c.} \right) \,.
\end{eqnarray}
We expand the $\Gamma$ functions as before in $\omega$ and use
\begin{eqnarray}
\left(\frac{M_c}{\alpha}\right)^{\frac{i}{2}\omega}
 &=&
 e^{i(-\pi + \phi)}
 \;=\; - (\cos{\phi} + i\sin{\phi}) \,,
\end{eqnarray}
which we have just derived, to get
\begin{eqnarray}
\frac{\theta_c}{\alpha} &\simeq& \mp
\frac{448}{75} \frac{\Gamma(1/2)}{\Gamma(1/4)^2}
\left(\frac{M_c}{\alpha}\right)^{3/2}  \,\,,
\end{eqnarray}
to leading order in $\omega$, using the expression for
$\phi$, Eq.~\ref{eqappphics}.

\section{Stationary Point of the Effective Potential}

The effective potential given in section \ref{CJTefpot} reproduces the
SD equation for $B_{\pm}$ at the stationary point as follows. The
variation with respect to $B_{\pm}$ of the effective potential is
calculated as follows:
\begin{eqnarray}
 {\delta \over \delta B_{\pm}(p)} V_0[B]
&=& {\delta \over \delta B_{\pm}(p)}
\int {{\rm d}^3p \over (2\pi)^3} \left[ - \ln \left( 1 +
{B_{\pm}^2(p) \over p^2} \right) + 2{B_{\pm}^2(p) \over
p^2+B_{\pm}^2(p)} \right]
\nonumber\\ &=&   {1 \over (2\pi)^3} 2B_{\pm}(p)
 {[p^2-B_{\pm}^2(p)] \over [p^2+B_{\pm}^2(p)]^2}\,,
\end{eqnarray}
and
\begin{eqnarray}
  {\delta \over \delta B_{\pm}(p)} V_1[B]
 &=&  -{2e^2 \over  (2\pi)^3}
 {[p^2-B_{\pm}^2(p)] \over [p^2+B_{\pm}^2(p)]^2}
 \int {{\rm d}^3k \over (2\pi)^3}
 {1 \over k^2+B_{\pm}^2(k)}
\nonumber\\&&
\times
\left[ \left( 2\, D_T(q) + \frac{a(q)}{q^2}\right) B_{\pm}(k)
\mp 2 {(k\cdot q) \over |q|} D_O(q) \right] \,.
\end{eqnarray}
Therefore we obtain
\begin{eqnarray}
 {\delta \over \delta B_{\pm}(p)} V[B]
&=&  {2 \over (2\pi)^3}
 {[p^2-B_{\pm}^2(p)] \over [p^2+B_{\pm}^2(p)]^2}
  \Biggr\{ B_{\pm}(p) -  e^2 \int {{\rm d}^3k \over (2\pi)^3}
 {1 \over k^2+B_{\pm}^2(k)}
\nonumber\\&&
\times
\left[ \left( 2\, D_T(q) + \frac{a(q)}{q^2}\right) B_{\pm}(k)
\mp 2 {(k\cdot q) \over |q|} D_O(q) \right]
\Biggr\} \,.
\end{eqnarray}
This shows that the SD equation for $B_{\pm}$ is obtained from the
stationary condition $\delta V[B]/\delta B_{\pm}=0$ of the CJT
effective potential $V[B]$.

%
%
\begin{figure}
%
\caption{
Scaling of the fermion mass and the chiral condensates:
the quantities $\omega\,\ln(m(0)/\alpha)$ and
$\omega\,\ln(\langle\bar\psi\psi\rangle/\alpha^2)$ as functions of
$\omega = \protect{\sqrt{N_c/N - 1}}$ for the linearized equation
(analytical solution, dashed line) and the full nonlinear integral
equation (numerically, solid line) in pure QED.}
\label{fig1}
\bigskip
%
%
\caption{
The fermion mass function $m(p)$ in pure QED: the analytical solution
(dashed line) and numerical one (solid line) for $N = 2$ and $N = 3$
where both the horizontal and the vertical axes are logarithmic. }
\label{fig2}
\bigskip
%
%
\caption{
The numerical solutions for wave-function renormalization functions,
$A_+(p)$ and $\tilde A_-(p)$, and mass functions, $B_+(p)$, $\tilde
B_-(p)$, $B_e(p)$, $B_o(p)$ at $\theta = 0.4\cdot10^{-5}$ and $m(p)$
at $\theta=0$ for $N = 3$.  (a) $1 - A_+(p)$, $1 - \tilde A_-(p)$, (b)
mass functions in the IR region, (c) the same as (b) in the
intermediate region.  }
\label{fig3}
\bigskip
%
%
\caption{
The infrared values $B_+(0)$ (dashed line) and $\tilde B_-(0)$
(solid line) as functions of $N$ for some different values of
$\theta$, obtained numerically by solving the two sets of coupled
integral equations. The dotted line indicates $m(0)$ in the case
of $\theta=0$.}
\label{fig4}
\bigskip
%
%
\caption{
The infrared values for $B_+(0)$, $\tilde B_-(0)$, $B_e(0)$, $B_o(0)$,
and the chiral condensate $\langle \bar \psi \psi \rangle$ as
functions of $\theta$ for $N=3$, obtained numerically by solving the
two sets of coupled integral equations. Note that the scale for the
condensate is different.}
\label{fig5}
\bigskip
%
%
\caption{
The infrared values $B_+(0)$ (dashed lines) and $\tilde B_-(0)$
(solid lines) as obtained analytically: (a) as function of $\theta$
for $N = 3$, rescaled and compared with our numerical solutions,
and (b) as function of $N$ for some different values of $\theta$,
which should be compared with the numerical results plotted in
Fig.~\protect{\ref{fig4}}.}
\label{fig6}
\bigskip
%
%
\caption{
Phase diagram in the $(N,\theta)$-plane: the critical line for the
chiral phase transition obtained analytically and some numerical
estimates for the critical point.}
\label{fig7}
\bigskip
%
%
\caption{
The chiral condensate and $B_e(0)$ as functions of $N$ for
$\theta = 10^{-5}$ and for pure QED ($\theta=0$), obtained numerically
and analytically. Note that the scale for the condensate and $B_e(0)$
is different.}
\label{fig8}
\bigskip
%
%
\caption{
The analytical solution of the linearized equation and the
numerical solution of the full integral equation for $N = 3$
and $\theta = 0.4\cdot 10^{-5}$.}
\label{fig9}
%
\end{figure}


\begin{thebibliography}{99}
\bibitem[*]{emkon}
 e-mail: kondo@cuphd.nd.chiba-u.ac.jp
\bibitem[\dagger]{emmar}
 e-mail: maris@eken.phys.nagoya-u.ac.jp

\bibitem{KM94}
  K.-I. Kondo and P. Maris,
  {\it First-order Phase Transition in Three-dimensional
  QED with Chern-Simons Term},
  Chiba/Nagoya Univ. Preprint, CHIBA-EP-84/DPNU-94-33,
  (hep-ph/9408210), to be published in Phys. Rev. Lett.

\bibitem{ABKW86}
  T. Appelquist, M. Bowick, D. Karabali and L.C.R.
  Wijewardhana,
  Phys. Rev. D 33 (1986) 3704.

\bibitem{Pisarski84}
  R.D. Pisarski,
  Phys. Rev. {\bf D29} (1984) 2423.

\bibitem{SW88}
  G.W. Semenoff and L.C.R. Wijewardhana,
  Phys. Rev. Lett. {\bf 62} (1988)  2633.

\bibitem{Miransky85}
  V.A. Miransky,
  Nuovo Cimento 90A (1985) 149.

\bibitem{DJT81}
  S. Deser, R. Jackiw and S. Templeton,
  Ann. Phys. 140 (1981) 372.

\bibitem{Fradkin91}
  E. Fradkin,
  {\it Field Theories of Condensed Matter Systems},
  (Addison-Wesley Publishing Company, 1991).

\bibitem{Wilczek90}
  F. Wilczek,
  {\it Fractional Statistics and Anyon Superconductivity},
  (World Scientific, 1990)

\bibitem{DM92}
  N. Dorey and N.E. Mavromatos,
  Nucl. Phys. B 386 (1992) 614.

\bibitem{ABKW86b}
  T. Appelquist, M.J. Bowick, D. Karabali and L.C.R.
  Wijewrdhana,
  Phys. Rev. D 33 (1986) 3774.

\bibitem{Poly88}
  A.P. Polychronakos,
  Phys. Rev. Lett. 60 (1988) 1920.

\bibitem{RY86a}
  S. Rao and R. Yahalom,
  Phys. Lett. B 172 (1986) 227;
  Phys. Rev. D 34 (1986) 1194.

\bibitem{HM89}
  Y. Hoshino and T. Matsuyama,
  Phys. Lett. B 222 (1989) 493.

\bibitem{VW84}
  C. Vafa and E. Witten,
  Nucl. Phys. B 234 (1984) 173.

\bibitem{CCW91}
  M. Carena, T.E. Clark and C.E.M. Wagner,
  Nucl. Phys. B 356 (1991) 117;
  Phys. Lett. B 259 (1991) 128.

\bibitem{KEIT94}
  K.-I. Kondo, T. Ebihara, T. Iizuka and E. Tanaka,
  {\it Dynamical Breakdown of Chirality and Parity in (2+1)-dimensional QED},
  Chiba Univ. Preprint, CHIBA-EP-77-REV,
  (hep-ph/9404361, revised), to be published in Nucl. Phys. B.

\bibitem{RW94}
  For a review, see
  C.D. Roberts and A.G. Williams,
  Progr. Part. and Nucl. Phys. {\bf 33} (1994) 477,
  and the
  Proceedings of the 1991 Nagoya Spring School  on Dynamical
  Symmetry Breaking, K. Yamawaki (editor),
  (World Scientific, Singapore, 1992).

\bibitem{KN89}
  K.-I. Kondo and H. Nakatani,
  Mod. Phys. Lett. {\bf A4} (1989) 2155.

\bibitem{KN90}
  K.-I. Kondo and H. Nakatani,
  Mod. Phys. Lett. A 5 (1990) 407.

\bibitem{DKK89}
  E. Dagotto, J.B. Kogut and A. Koci\'c,
  Phys. Rev. Lett. 62 (1989) 1083-1086; E. Dagotto, A.
Koci\'c and J.B. Kogut,
  Nucl. Phys. B 334 (1990) 279.

\bibitem{CPW92}
  D.C. Curtis, M.R. Pennington and D. Walsh,
  Phys. Lett. B 295 (1992) 313;
  M.R. Pennington and D. Walsh,
  Phys. Lett. B 253 (1991) 246.

\bibitem{AJM90}
  D. Atkinson, P.W. Johnson and P. Maris,
  Phys. Rev. D 42 (1990) 602.

\bibitem{CP91}
  D.C. Curtis and M.R. Pennington,
  Phys. Rev. D 44 (1991) 536.

\bibitem{ABGPR93}
  D. Atkinson, V.P. Gusynin and P. Maris,
  Phys. Lett. B 303 (1993) 157;
  D. Atkinson, J.C.R. Bloch, V.P. Gusynin, M.R. Pennington
  and M. Reenders,
  Phys. Lett. B 329 (1994) 117;

\bibitem{Kondo92}
  K.-I. Kondo, Intern. J. Mod. Phys. {\bf A 7} (1992) 7239.

\bibitem{JT81}
  R. Jackiw and S. Templeton,  Phys. Rev.
  {\bf D23} (1981) 2291;
  T.W. Appelquist and R.D. Pisarski,
  Phys. Rev. {\bf D23} (1981)  2305;
  T.W. Appelquist and U. Heinz,
  Phys. Rev. {\bf D23} (1981) 2169.

\bibitem{ANW88}
  T. Appelquist, D. Nash and L.C.R. Wijewardhana,
  Phys. Rev. Lett. 60 (1988) 2575.

\bibitem{Nash89}
  D. Nash,
  Phys. Rev. Lett. 62 (1989) 3024.

\bibitem{PW88}
  M.R. Pennington and S.P. Webb,
  {\it Hierarchy of scales in three dimensional QED},
  BNL-40886, January 1988 (unpublished).

\bibitem{AJP88}
  D. Atkinson, P.W. Johnson and M.R. Pennington,
  {\it Dynamical mass generation in three-dimensional QED},
  BNL-41615, August 1988 (unpublished).

\bibitem{Azcoiti93}
  V. Azcoiti, X.-Q. Luo, C.E. Piedratita, G. Di Carlo, A.F. Grillo  and A.
Galante,
  Phys. Lett. B 313 (1993) 180.

\bibitem{Nakatani88}
  H. Nakatani,
  Comment given at the 1988 International Workshop on New
  Trend in Strong Coupling Gauge Theories, Nagoya, Aug.
  24-27 (unpublished).

\bibitem{GSC90}
  H. Georgi, E.H. Simmons and A.G. Cohen,
  Phys. Lett. B 236 (1990) 183;
  E.H. Simmons,
  {\it Comment on higher-order corrections in
  (2+1)-dimensional
  QED},
  NSF-ITP-90-26, HUTP-90/A009.

\bibitem{KM92}
  T. Kugo and M.G. Mitchard,
  Phys. Lett. B 282 (1992) 162.

\bibitem{KN92}
  K.-I. Kondo and H. Nakatani,
  Prog. Theor. Phys. 87 (1992) 193.

\bibitem{CJT74}
  J. Cornwall, R. Jackiw and E. Tomboulis,
  Phys. Rev. D 10 (1974) 2428.

\bibitem{Atkinson}
  D. Atkinson,
  J. Math. Phys. 28 (1987) 2494;
  D. Atkinson, P.W. Johnson and K. Stam,
  Phys. Lett. B 201 (1988) 105.

\bibitem{Kondo92p}
  K.-I. Kondo,
  {\it Triviality Problem and Schwinger-Dyson Equation
Approach,
  in Dynamical Symmetry Breaking},  ed. K. Yamawaki (World
  Scientific Pub. Co.,  Singapore, 1992).

\bibitem{thes}
  P. Maris,
 {\it Nonperturbative analysis of the fermion propagator:
  complex singularities and dynamical mass generation},
  Ph.D. Thesis, U. of Groningen, October 1993.

\bibitem{HP93}
  D.K. Hong and S.H. Park,
  Phys. Rev. D47 (1993) 3651.

\bibitem{Haymaker91}
  R.W. Haymaker,
  Rivista del Nuovo Cimento, 14 (1991) No. 8, 1.

\bibitem{Matsuki91} T. Matsuki,
  Z. Phys. C51 (1991) 429;
  {\it Stability at the origin in (2+1)-dimensional QED},
  in 1989 Workshop on Dynamical Symmetry Breaking, Nagoya
  University, 1989.

\bibitem{MMV88}
  T. Matsuki, L. Miao and K.S. Viswanathan, Simon Fraser
  Univ. Preprint, June 1987 (revised: May 1988).

\bibitem{Creutz83}
  See for example, M. Creutz,
  {\it Quarks, Gluons and Lattices},
  (Cambridge University Press, 1983);
  E. Seiler,
  {\it Gauge Theories as a Problem of Constructive Quantum Field Theory and
   Statistical Mechanics}, Lecture Notes in Physics 159
  (Springer Verlag, 1982).

\bibitem{lk55} L.D. Landau and I.M. Khalatnikov,  J. Exper. Theor. Phys.
  USSR {\bf 29}, 89, (1955),  (translation: Sov. Phys. JETP {\bf 2}, 69
  (1956)).


\end{thebibliography}
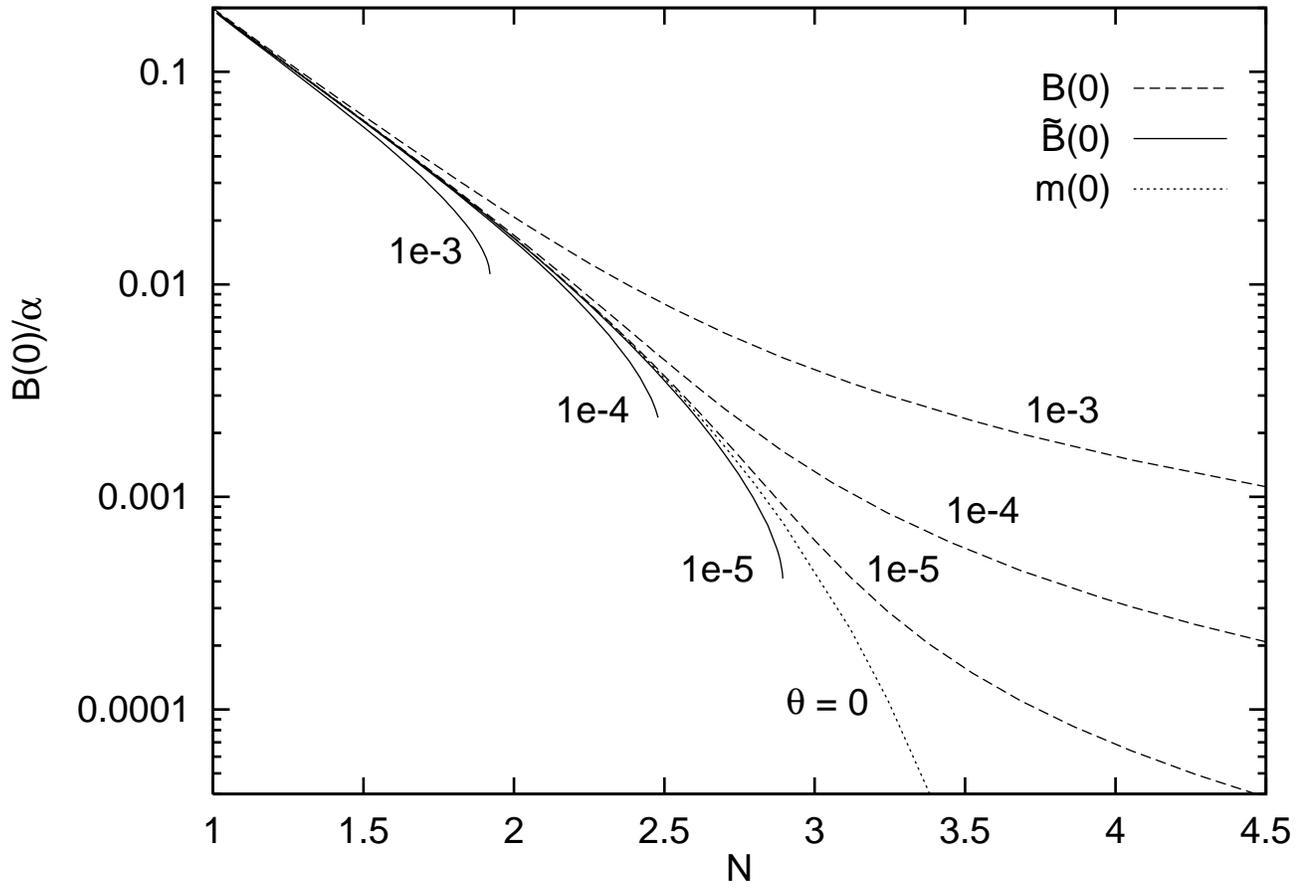
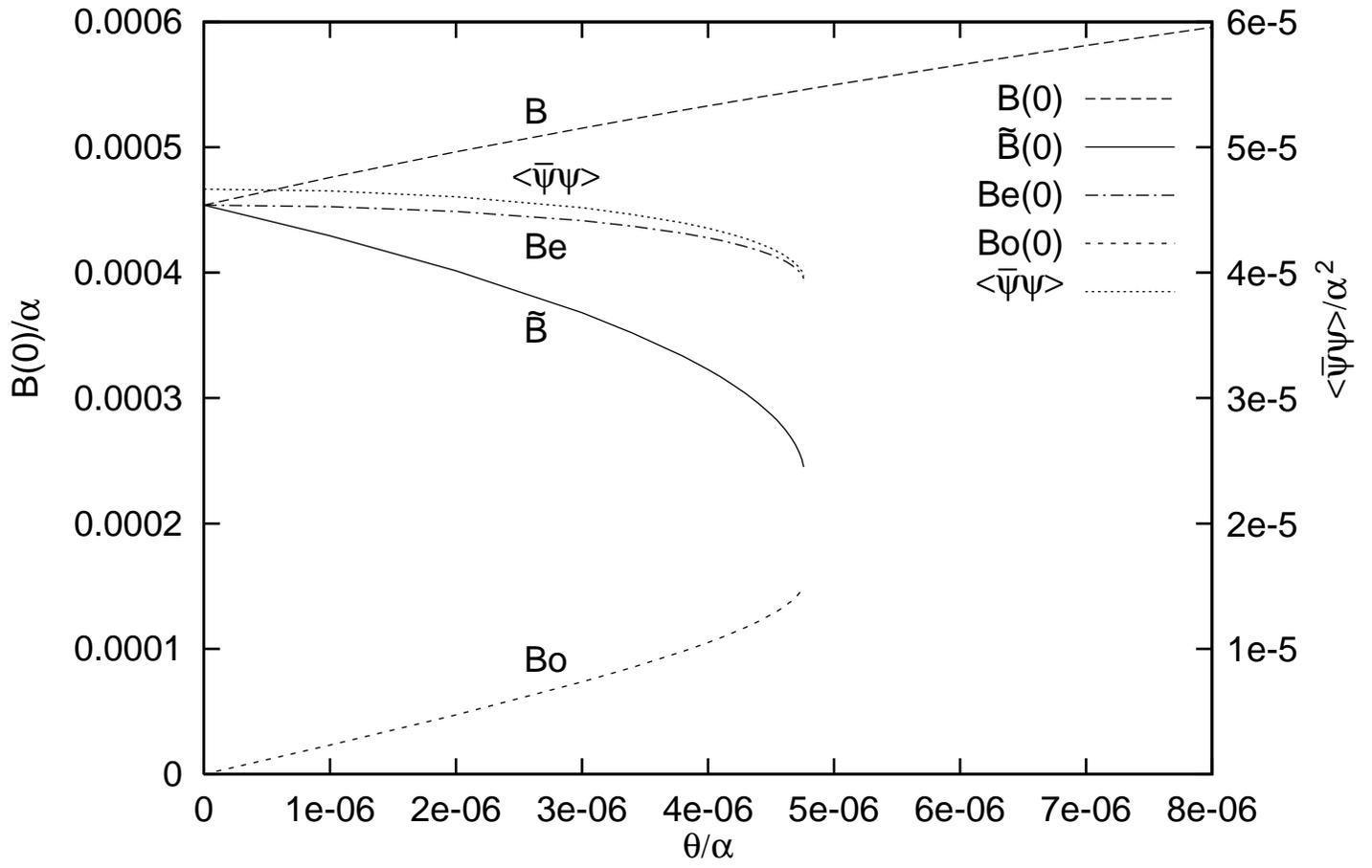
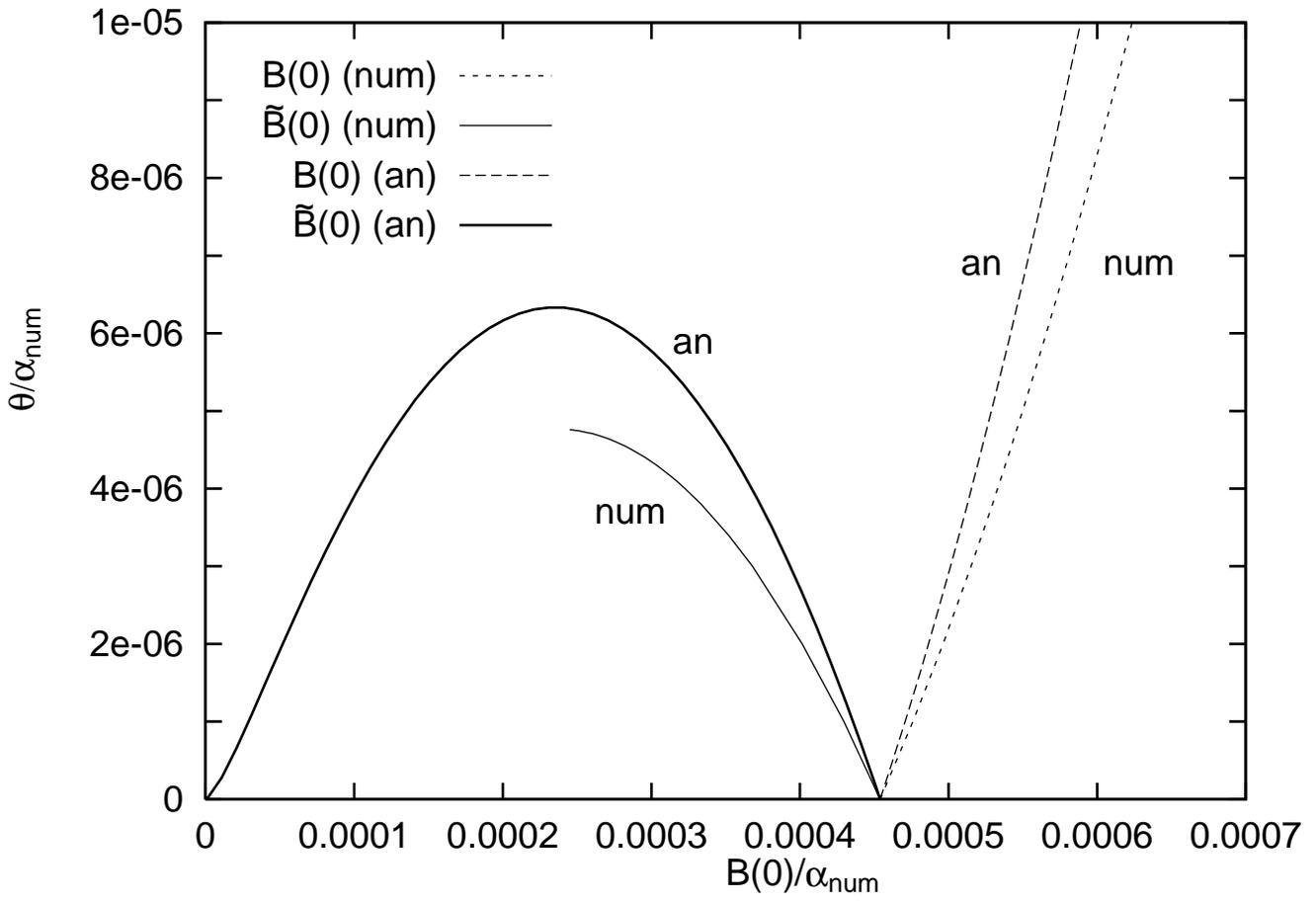
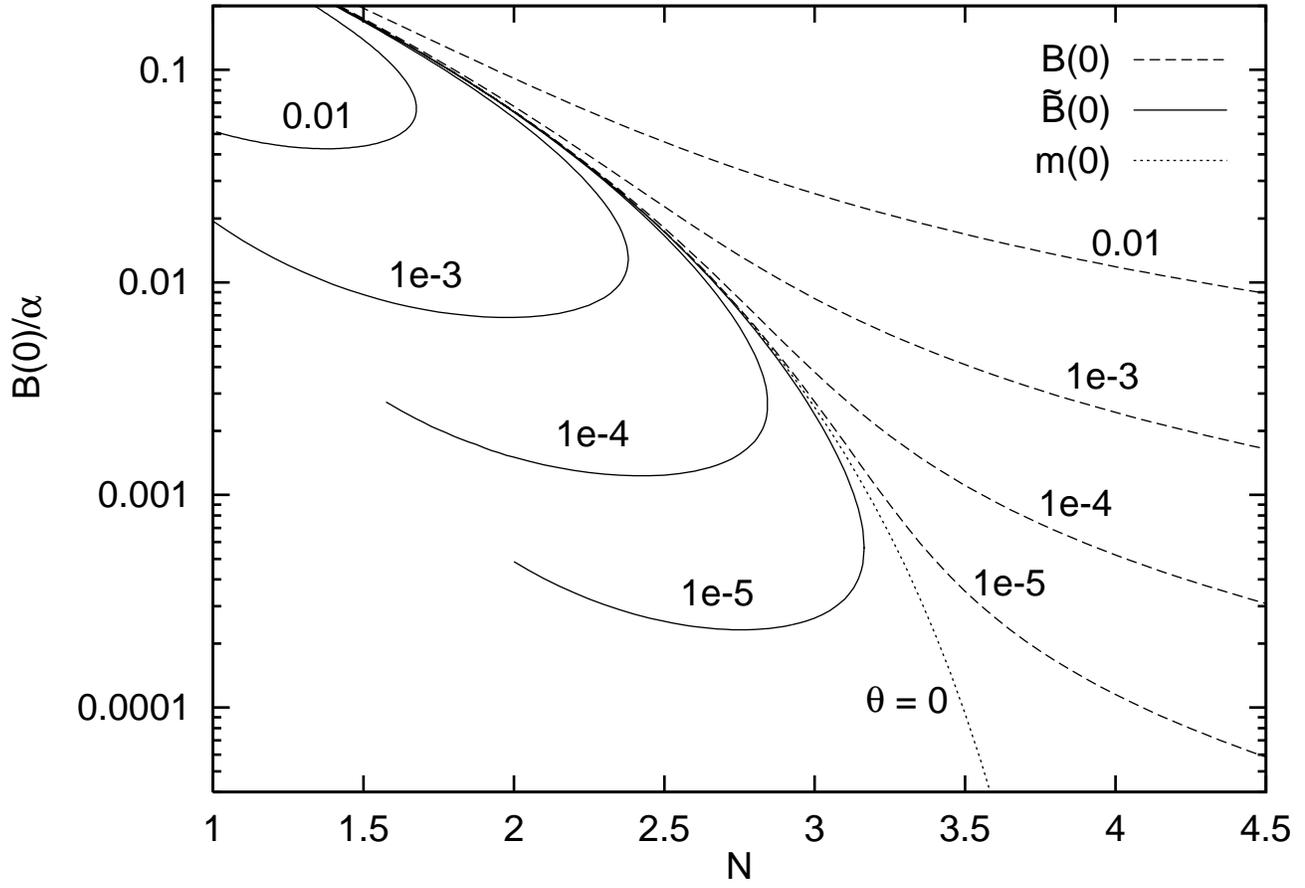
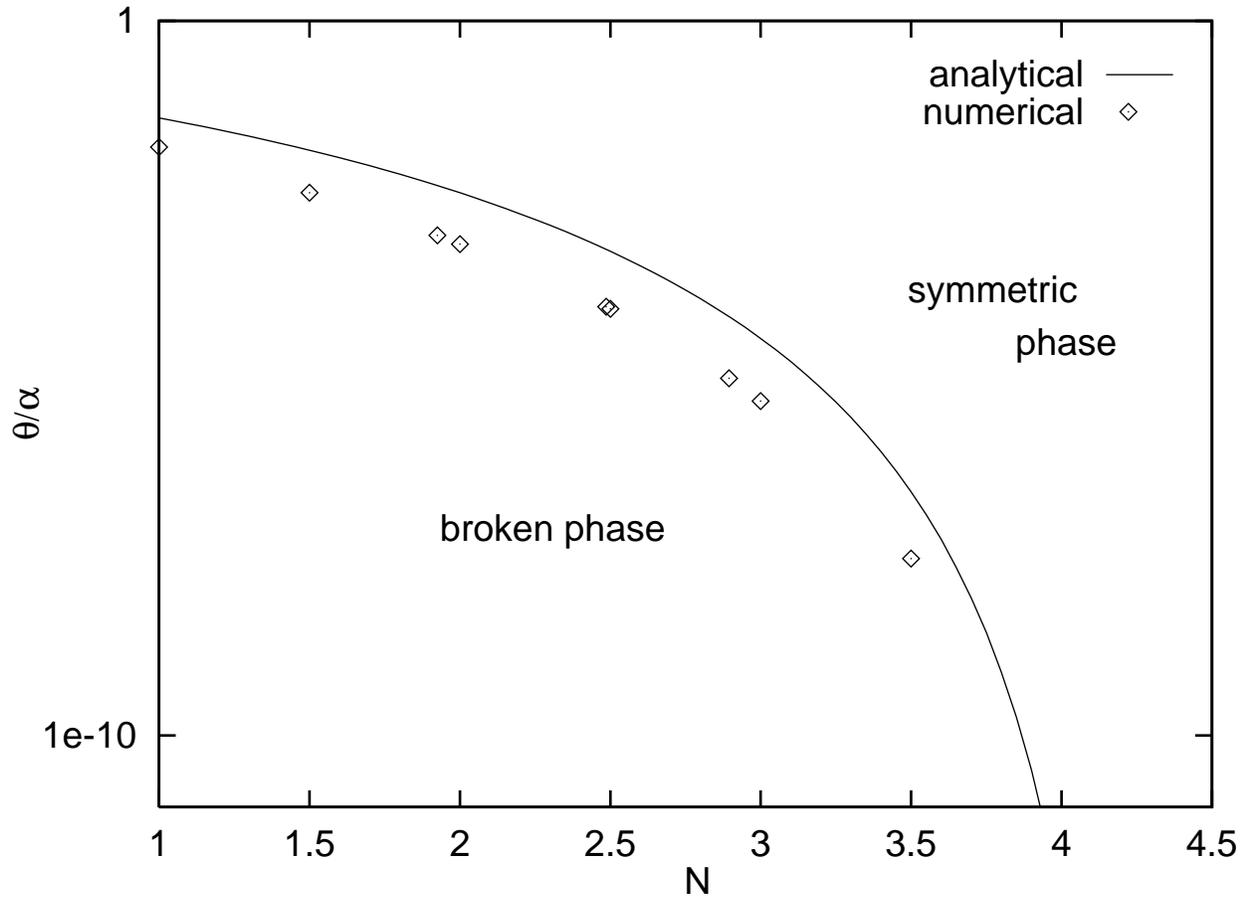
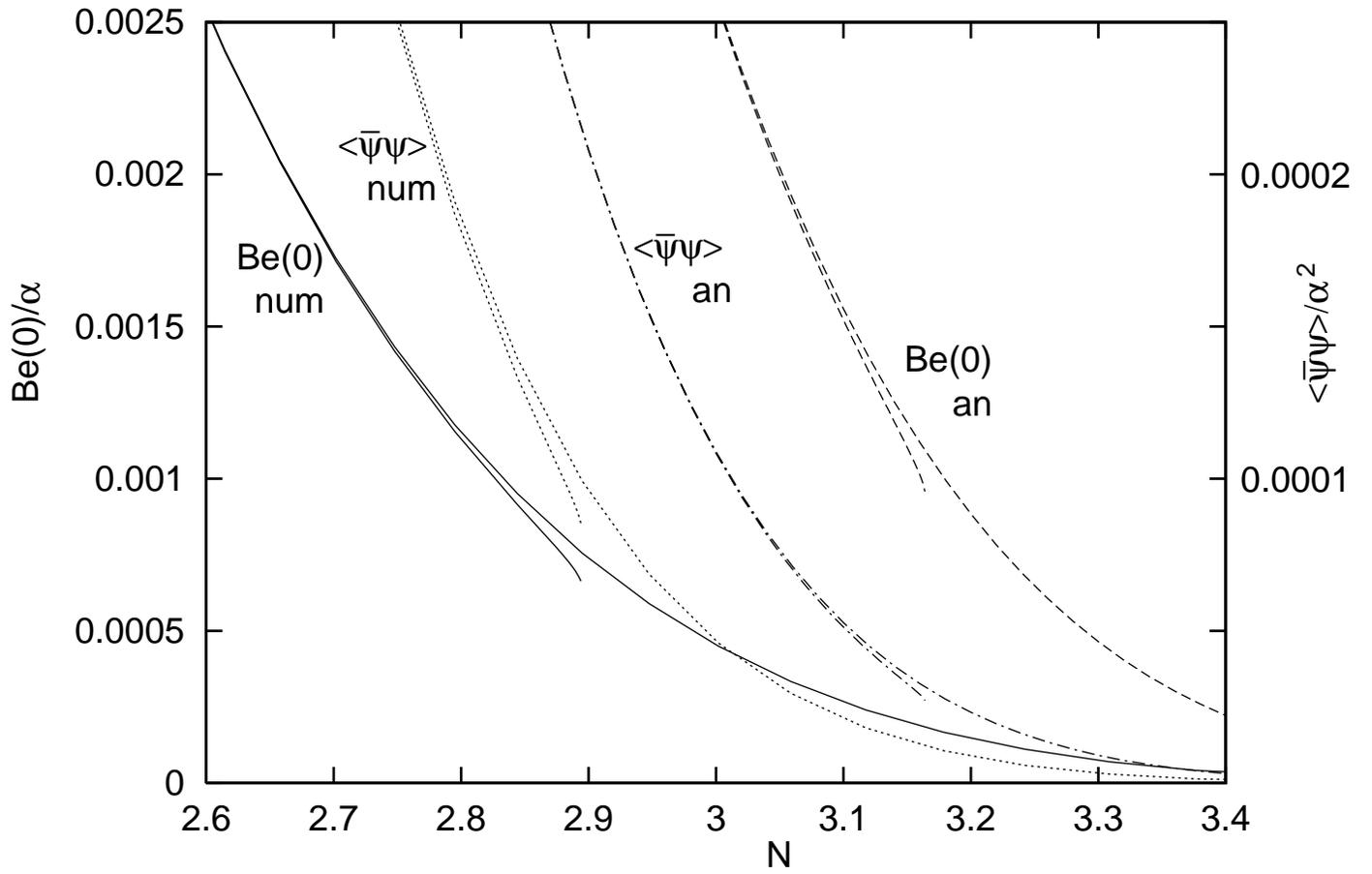
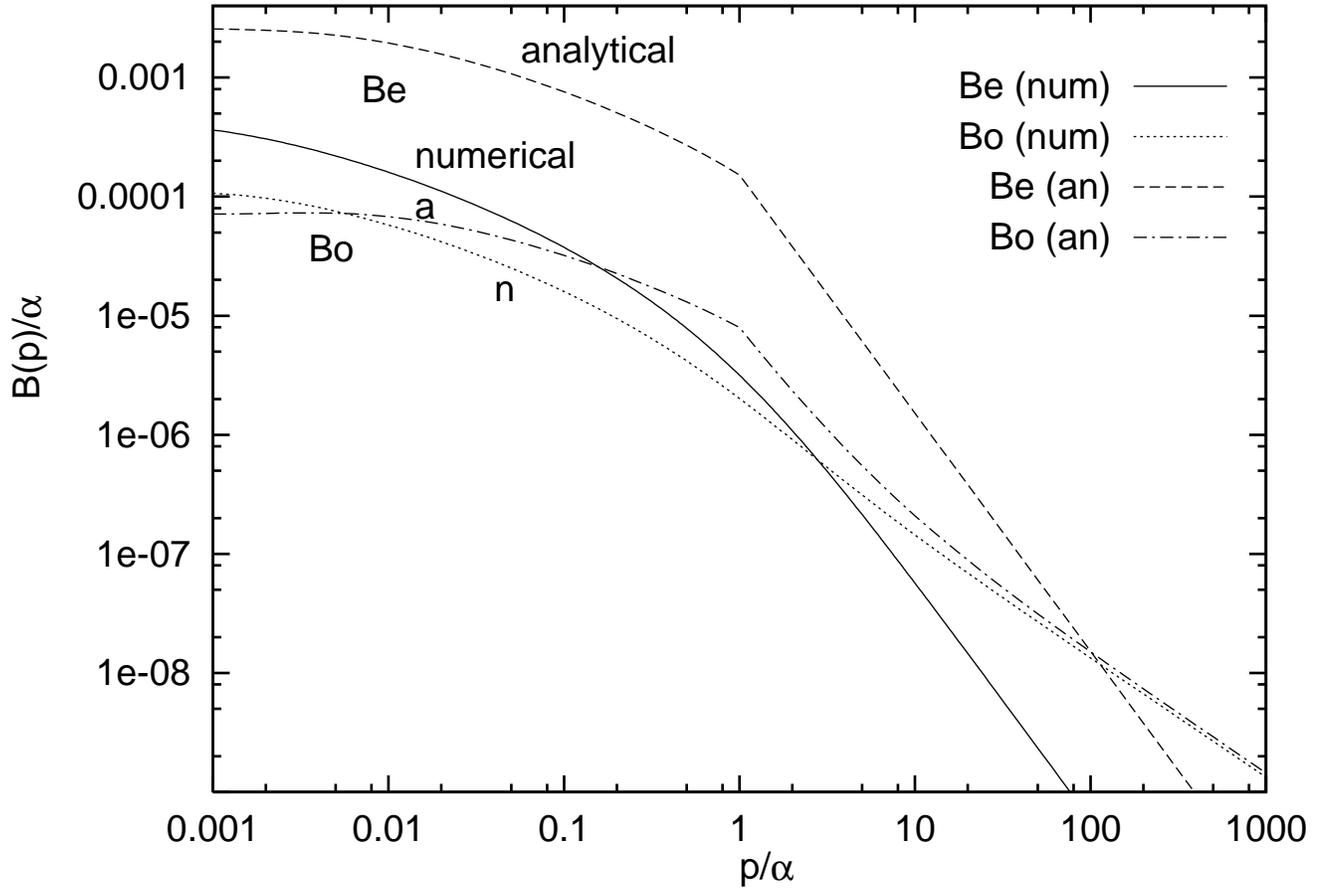
\end{document}